\documentclass[aps,floatfix,showpacs,showkeys,superscriptaddress,preprint]{revtex4}

\usepackage{latexsym,graphicx,epsfig,psfrag}

\usepackage{epstopdf}
\usepackage{amssymb,amsmath,amsxtra,amsfonts}
\usepackage{bm}

\usepackage{longtable}
\usepackage{multirow}
\usepackage{booktabs}
\usepackage{array}
\usepackage{wrapfig}
\usepackage{color}
\usepackage{titlesec}

\titleformat{\section}{\large\bfseries}{\thesection}{1em}{}

\newcommand{\bea}{\begin{eqnarray}}
\newcommand{\ena}{\end{eqnarray}}
\newcommand{\be}{\begin{equation}}
\newcommand{\en}{\end{equation}}
\newcommand{\nn}{\nonumber\\}
\newcommand{\ed}{\end{document}} 

\newcommand{\ord}{\mathcal{O}}

\newcommand{\la}{\langle}
\newcommand{\ra}{\rangle}

\newcommand{\re}{{\rm Re}}

\newcommand{\Hs}{{\hat s}}
\newcommand{\Hx}{{\hat x}}

\newcommand{\Ht}{{\hat t}}
\newcommand{\Hu}{{\hat u}}
\newcommand{\Hml}{{\hat m_\ell}}
\newcommand{\Hmb}{{\hat m_b}}
\newcommand{\HfBs}{{\hat f_{B_s}}}

\begin{document}

\title{Study of \boldmath{$B_s\to \ell^+\ell^-\gamma$} decays in covariant quark model} 

\author{S. Dubni\v{c}ka}
\affiliation{Institute of Physics, Slovak Academy of Sciences, D\'{u}bravsk\'{a} cesta 9, 84511 
Bratislava, Slovakia}

\author{A. Z. Dubni\v{c}kov\'a}
\affiliation{Department of Theoretical Physics, Comenius University, Mlynsk\'{a} dolina F1, 84248 
Bratislava, Slovakia}

\author{M. A. Ivanov}
\email{ivanovm@theor.jinr.ru}
\affiliation{Bogoliubov Laboratory of Theoretical Physics, 
Joint Institute for Nuclear Research, 141980 Dubna, Russia}

\author{A. Liptaj}
\email{andrej.liptaj@savba.sk}
\affiliation{Institute of Physics, Slovak Academy of Sciences, D\'{u}bravsk\'{a} cesta 9, 84511 
Bratislava, Slovakia}

\author{P. Santorelli}
\email{Pietro.Santorelli@na.infn.it}
\affiliation{Dipartimento di Fisica ``E.~Pancini'', Universit\`{a} di Napoli Federico II, Complesso Universitario di Monte S. Angelo, Via Cintia, Edificio 6, 80126 Napoli, Italy}
\affiliation{Istituto Nazionale di Fisica Nucleare, Sezione di Napoli, 80126 Napoli, Italy}

\author{C. T. Tran}
\email{tran@na.infn.it}
\thanks{corresponding author}
\affiliation{Dipartimento di Fisica ``E.~Pancini'', Universit\`{a} di Napoli Federico II, Complesso Universitario di Monte S. Angelo, Via Cintia, Edificio 6, 80126 Napoli, Italy}
\affiliation{Istituto Nazionale di Fisica Nucleare, Sezione di Napoli, 80126 Napoli, Italy}

\date{\today}

\begin{abstract}
We study the rare radiative leptonic decays  $B_s\to \ell^+\ell^-\gamma$ ($\ell=e,\mu,\tau$)
within the Standard Model, considering both the structure-dependent amplitude and bremsstrahlung. In the framework of the covariant confined quark model developed by us, we calculate the form factors characterizing the $B_s\to\gamma$ transition in the full kinematical region of the dilepton momentum squared
and discuss their behavior. We provide the analytic formula for the differential decay distribution and give predictions for
the branching fractions in both cases: with and without long-distance contributions. Finally, we compare our results with those obtained in other approaches.
  
\end{abstract}


\maketitle

\section{Introduction}
\label{sec:intro}
The rare radiative decays $B_q\to\ell^+\ell^-\gamma$ with $\ell=e,\mu,\tau$
and $q=d,s$ are of great interest for several reasons.  First, they are 
complementary to the well-known decays $B\to K^{(\ast)}\ell^+\ell^-$ and therefore provide us with extra tests of the Standard Model (SM) predictions
for processes which proceed at loop level.  Second, this process is 
not helicity suppressed as compared with the pure leptonic decays
$B_q\to\ell^+\ell^-$ due to the appearance of a photon in the final state. Theoretical estimates of the decay branching fractions
have shown that ${\cal B}\left(B_s\to \mu^+\mu^-\gamma\right)$ may be an order
of magnitude larger than ${\cal B}\left(B_s\to \mu^+\mu^-\right)$.  

There are a number of theoretical calculations of the branching fractions
${\cal B}(B_q\to\ell^+\ell^-\gamma)$ performed in different approaches. 
Among them one can mention the early studies in the framework of
a constituent quark model~\cite{Eilam:1996vg}, light-cone QCD sum 
rules~\cite{Aliev:1996ud,Aliev:1997sd}, and the light-front 
model~\cite{Geng:2000fs}.
The structure-dependent amplitude of the decays $B_s\to\ell^+\ell^-\gamma$ 
was analyzed in Ref.~\cite{Dincer:2001hu} by taking a universal form 
for the form factors, which is motivated by QCD and related to the 
light-cone wave
function of the $B_s$ meson. 
In Ref.~\cite{Kruger:2002gf} it was shown that efficient constraints on the behavior of the form factors can be obtained from the gauge-invariance requirement of the $B_q\to\ell^+\ell^-\gamma$ amplitude, as well as from the resonance structure of the form factors and their relations at large photon energies.  
Universality of nonperturbative QCD effects in radiative B decays
was studied in Ref.~\cite{DescotesGenon:2002ja}. 
In Ref.~\cite{Melikhov:2004mk} long-distance QCD effects in  the $B_{d/s}\to\ell^+\ell^-\gamma$ decays were 
analyzed. It was shown that the contribution
of light vector-meson resonances related to the virtual photon emission 
from valence quarks of the $B$ meson gives a sizable impact
on the dilepton differential distribution. 
In Ref.~\cite{Kozachuk:2017mdk} the $B_{d/s}\to\gamma$ transition form factors were calculated
within the relativistic dispersion approach based on the constituent quark 
picture. A detailed analysis of the charm-loop 
contributions to the radiative leptonic decays was also performed.
Very recently, a novel strategy to search for  the decays 
$B_s\to\mu^+\mu^-\gamma$  
in the event sample selected for  $B_s\to\mu^+\mu^-$ searches was 
presented~\cite{Dettori:2016zff}.

It is worth noting that the predictions for the branching fractions $\mathcal{B}(B_s\to\ell^+\ell^-\gamma)$ given in the literature are still largely different from each other, ranging from $2.4\times 10^{-9}$ to $2.5\times 10^{-8}$ for the electron mode, and from $1.9\times 10^{-9}$ to $1.9\times 10^{-8}$ for the muon one~\cite{Aliev:1996ud, Melikhov:2004mk}. Moreover, in some early calculations, the long-distance contributions from the $c\bar{c}$ resonances were neglected~\cite{Eilam:1996vg, Dincer:2001hu}. Note that in Refs.~\cite{Eilam:1996vg, Aliev:1996ud, Geng:2000fs} the authors concluded that the contributions from diagrams with the virtual photon emitted from the valence quarks of the $B_s$ meson are small and therefore, neglected them. However, as we will show later on, the diagram with the virtual photon emission from the light $s$ quark gives a very sizable contribution. 

Putting aside the total branching fraction, the shape of the hadronic form factors are important. In particular, it directly affects the decay distribution, and therefore the partial branching fraction integrated in different $q^2$ bins, which is more important for experimental studies than the total branching fraction. Regarding the form factor shape, the authors of Refs.~\cite{Dincer:2001hu, Kruger:2002gf} have pointed out that the form factors $F_{TV}$ and $F_{TA}$ in Ref.~\cite{Aliev:1996ud} may be unreliable since they strongly violate the relation $F_{TV}\approx F_{TA}$ at large photon energies. Also, the form factors $F_{TV}$ and $F_{TA}$ obtained in Ref.~\cite{Geng:2000fs} vanish at maximum transferred momentum $q^2=m_{B_s}^2$, which seems unrealistic~\cite{Kruger:2002gf}. Among the model-based approaches, the most reliable form factors in the whole $q^2$ range are provided in Refs.~\cite{Dincer:2001hu, Kozachuk:2017mdk}. However, in Ref.~\cite{Dincer:2001hu}, the resonances were not taken into account. Note that the light resonance $\phi$ is important since it
significantly enhances the partial branching fraction in the low $q^2$ region, which is the main source of the signal for these decays at the LHC~\cite{Melikhov:2004mk}.

In the literature there exist also model-independent studies of the $B_s\to\ell^+\ell^-\gamma$ and related decays $B_s\to\ell^+\ell^-$ and $B_s\to\ell^+\nu_\ell\gamma$~\cite{DescotesGenon:2002ja, Korchemsky:1999qb, Lunghi:2002ju, Bosch:2003fc, Burdman:1994ip}. However, most of them focus mainly on the form factors $F_V$ and $F_A$. Also, the form factors were given with high accuracy only in a limited kinematical range, usually the range where the photon energy $E_\gamma$ is much higher than the QCD scale. In Ref.~\cite{Aditya:2012im}, model-independent predictions for $\mathcal{B}(B_s\to\mu^+\mu^-\gamma)$ were provided, but only for the low-$E_\gamma$ region.

In this paper we calculate the matrix elements and the differential decay rates
of the decays $B_s\to\ell^+\ell^-\gamma$ in the framework of the covariant
confined quark model previously developed by us (see, e.g.,
Ref.~\cite{Branz:2009cd}). This is a quantum-field theoretical model
based on relativistic Lagrangians which effectively describe
the interaction of hadrons with their constituent quarks. 
The quark confinement is realized by cutting the integration variable,
which is called the proper time, at the upper limit. The interaction with 
the electromagnetic field is introduced by gauging the interaction
Lagrangian in such a way as to keep the gauge invariance of the matrix elements
at all calculation steps. This model has been successfully applied for
the description of the matrix elements and form factors in the full kinematical
region in semileptonic and rare decays of heavy mesons
as well as baryons (see, e.g., Refs.~\cite{Ivanov:2011aa,Faessler:2002ut,Dubnicka:2015iwg, Gutsche:2015mxa,Gutsche:2013pp}).
 
The rest of the paper is organized as follows. In Sec.~\ref{sec:model}
we give a necessary brief sketch of our approach. 
The introduction of electromagnetic interactions in the model is described in Sec.~\ref{sec:em}. Section~\ref{sec:ff} is devoted to the calculation
of the decay matrix elements.
We also briefly discuss their gauge invariance.
In Sec.~\ref{sec:dis} we recalculate the formula for
the twofold decay distribution in terms of the Mandelstam variables $(t,s)$. Then we integrate out
the $t$ variable analytically and present the expression for the dilepton
differential distribution. In Sec.~\ref{sec:result} we provide numerical results for the form factors, the differential decay widths, and the branching fractions. 
A comparison with existing results in the literature is included. Finally, we conclude in Sec.~\ref{sec:summary}.


\section{Brief sketch of the covariant confined quark model}
\label{sec:model}
The covariant confined quark model (CCQM) has been developed by our group in a series of papers. In this section, we mention several key elements of the model only for completeness. For a more detailed description of the model, as well as the calculation techniques used for the quark-loop evaluation, we refer to Refs.~\cite{Branz:2009cd,Ivanov:2011aa,Faessler:2002ut,Gutsche:2018nks,Dubnicka:2015iwg, Gutsche:2015mxa,Gutsche:2013pp,Ivanov:2015woa} and references therein.

In the CCQM, the interaction Lagrangian of the $B_s$ meson with its constituent quarks is constructed from the hadron field $B_s (x)$ and
the interpolating quark current $J_{B_s}(x)$:
\be
{\mathcal{L}_{\rm int} = g_{B_s } B_s (x) J_{B_s}(x)+H.c.},
\label{eq:int_Lag}
\en
where the latter is given by
\be
J_{B_s}(x)  =  \int\!\! dx_1\!\!\int\!\! dx_2\: 
F_{B_s}(x;x_1,x_2)\bar b^a(x_1)i\gamma_5 s^a(x_2).
\en
The hadron-quark coupling $g_{B_s}$ is obtained with the help of the compositeness condition, which requires the wave function renormalization constant of the hadron to be equal to zero $Z_H=0$.
Here, $F_{B_s}(x;x_1,x_2)$ is the vertex function whose form is chosen so as to reflect the intuitive
expectations about the relative quark-hadron positions
\be
F_{B_s} (x;x_1,x_2) =  
\delta^{(4)} \left(x-w_1 x_1-w_2 x_2 \right)\:
\Phi_{B_s}\left[\left(x_1-x_2\right)^2\right],
\en
where we require $w_{1}+w_{2}=1$. We actually adopt the
most natural choice 
\be
w_1 = \frac{m_b}{m_b+m_s},\qquad 
w_2 = \frac{m_s}{m_b+m_s},
\en
in which the barycenter of the hadron is identified with that
of the quark system. The interaction strength 
$\Phi_{B_s}\left[\left(x_{1}-x_{2}\right)^{2}\right]$
is assumed to have a Gaussian form which is, in the momentum representation,
written as
\be
\widetilde{\Phi}_{B_s}\left(-p^2\right) =
\exp\left( p^2/\Lambda_{B_s}^2\right).
\en
Here, $\Lambda_{B_s}$ is a hadron-related size parameter, regarded as an
adjustable parameter of the model. 
 For the quark propagators $S_q$ we use the Fock-Schwinger representation
\be
S_q(k) = (m_{q}+\not\! k)
\int\limits_0^\infty\! d\alpha \exp[-\alpha(m_{q}^2-k^2)].
\en

Using various techniques described in our previous papers, a form factor $F$
can be finally written in the form of a threefold integral
\be
F   = \int\limits_0^{1/\lambda^2}\!\! dt\, t
\!\! \int\limits_0^1\!\! d\alpha_1
\!\! \int\limits_0^1\!\! d\alpha_2  \,
\delta\Big(1 -  \alpha_1-\alpha_2 \Big) 
f(t\alpha_1,t\alpha_2),
\label{eq:3fold}
\en
where $f(t\alpha_1,t\alpha_2)$ is the resulting integrand corresponding to
the form factor $F$, and $\lambda$ is the so-called infrared cutoff parameter,
which is introduced to avoid the appearance of the branching point
corresponding to the creation of free quarks, and taken to be universal
for all physical processes. The threefold integral in Eq.~(\ref{eq:3fold})
is calculated by using \textsc{fortran} code with the NAG library.

The model parameters are determined from a
least-squares fit to available experimental data and some lattice calculations.
We have observed that the errors of the fitted parameters are within 10$\%$. 
We calculated the propagation of these errors on the form factors and found
the uncertainties for the form factors to be of order 20$\%$ at small $q^2$
and 30$\%$ at high $q^2$~\cite{Soni:2018adu}.

In this paper we use the results of the updated fit
performed in Refs.~\cite{Gutsche:2015mxa,Ganbold:2014pua,Issadykov:2015iba}.
The central values of the model parameters 
involved in this paper are given by (in GeV)
\be
\begin{tabular}{ c c c c  c c  }
\quad $m_{u/d}$ \quad & \quad $m_s$ \quad & \quad $m_c$ \quad & \quad $m_b$ \quad
& \quad $\lambda$ \quad & \quad $\Lambda_{B_s}$ \quad  
\\\hline
 \quad 0.241 \quad & \quad 0.428 \quad & \quad 1.67 \quad &
\quad 5.04 \quad & \quad 0.181 \quad & \quad 2.05 \quad 
\end{tabular}
.
\label{eq:modelparam}
\en


\section{Electromagnetic interactions}
\label{sec:em}

Within the CCQM framework, interactions with electromagnetic fields are introduced as follows.
First, one gauges the free-quark Lagrangian 
in the standard manner by using minimal substitution
\be
\partial^\mu q_i \to (\partial^\mu - ie_{q_i} A^{\mu}) q_i
 \label{eq:em_min} 
\en 
that gives the quark-photon interaction Lagrangian 
\be 
{\cal L}^{\rm em-min}_{\rm int}(x) = 
\sum\limits_q  e_q  \bar q(x) \!\not\!\! A(x) q(x). 
\label{eq:Lagr_em_min}
\en 

In order to guarantee local invariance of the strong interaction 
Lagrangian, one multiplies 
each quark field $q(x_i)$ in ${\cal L}_{\rm int}^{\rm str}$ with a 
gauge field exponential.
One then has
\be
q_i(x_i)\to e^{-ie_{q_i} I(x_i,x,P)}  q_i(x_i),
\label{eq:gauging}
\en 
where
\be
I(x_i,x,P) = \int\limits_x^{x_i} dz_\mu A^\mu(z). 
\label{eq:path}
\en
The path $P$ connects the end points of the path integral.

It is readily seen that the full Lagrangian is invariant 
under the transformations
\bea
         q_i(x)  &\to &  e^{ie_{q_i} f(x)} q_i(x),\nn
 \bar{q}_i(x) &\to & \bar{q}_i(x) e^{-ie_{q_i} f(x)},\\
     A^\mu(x) &\to & A^\mu(x)+\partial^\mu f(x).\nonumber 
\label{eq:gauge_groop}
\ena

One then expands the gauge exponential
up to the required power of $e_{q}A_\mu$ needed in the perturbative series. 
This will give rise to a second term in the nonlocal electromagnetic 
interaction Lagrangian ${\cal L}^{\rm em-nonloc}_{\rm int}$. 
At first glance, it seems that the results will depend on the path $P$
taken to connect the end points of the path integral in 
Eq.~(\ref{eq:path}).  
However, one needs to know only the derivatives of the path integral
expressions when calculating the perturbative series.
Therefore, we use the 
formalism suggested in Refs.~\cite{Mandelstam:1962mi,Terning:1991yt}, 
which is based on the path-independent definition of the derivative of 
$I(x,y,P)$:
\be
\frac{\partial}{\partial x^\mu} I(x,y,P) = A_\mu(x).
\label{path2}
\en 

As a result of this rule, the Lagrangian describing the nonlocal interaction
of the $B_s$ meson, the quarks, and electromagnetic fields  reads (to the first order in the electromagnetic charge)

\bea
{\cal L}^{\rm em-nonloc}_{\rm int}(x) &=&
i g_{B_s} B_s(x) \int dx_1\int dx_2 \int dz 
\left( \bar b(x_1) \gamma_5 s(x_2) \right) A_\mu(z)
E^\mu(x;x_1,x_2,z),
\\
E^\mu(x;x_1,x_2,z) &=& 
\int\frac{d^4p_1}{(2\pi)^4}\int\frac{d^4p_2}{(2\pi)^4}
\int\frac{d^4q}{(2\pi)^4}
\exp[-ip_1(x_1-x)+ i p_2(x_2-x) + i q(z-x)]
\nn
&&
\times \Big\{
e_b (q^\mu w_2-2 p^\mu) w_2 \int\limits_0^1 d\tau
\tilde\Phi_{B_s}^\prime\left[-(p-w_2 q)^2\tau - p^2(1-\tau)\right]
\nn
&&
-e_s (q^\mu w_1+2 p^\mu) w_1 \int\limits_0^1 d\tau
\tilde\Phi_{B_s}^\prime\left[-(p+w_1 q)^2\tau - p^2(1-\tau)\right]
\Big\},
\label{eq:em-nonloc}
\ena
where $p=w_2p_1+w_1p_2$.
\section{Matrix elements of the decays \boldmath{$B_s\to \ell^+\ell^-\gamma$} }
\label{sec:ff}

\begin{figure*}[htbp]
\centering
\begin{tabular}{c}
\includegraphics[scale=0.5]{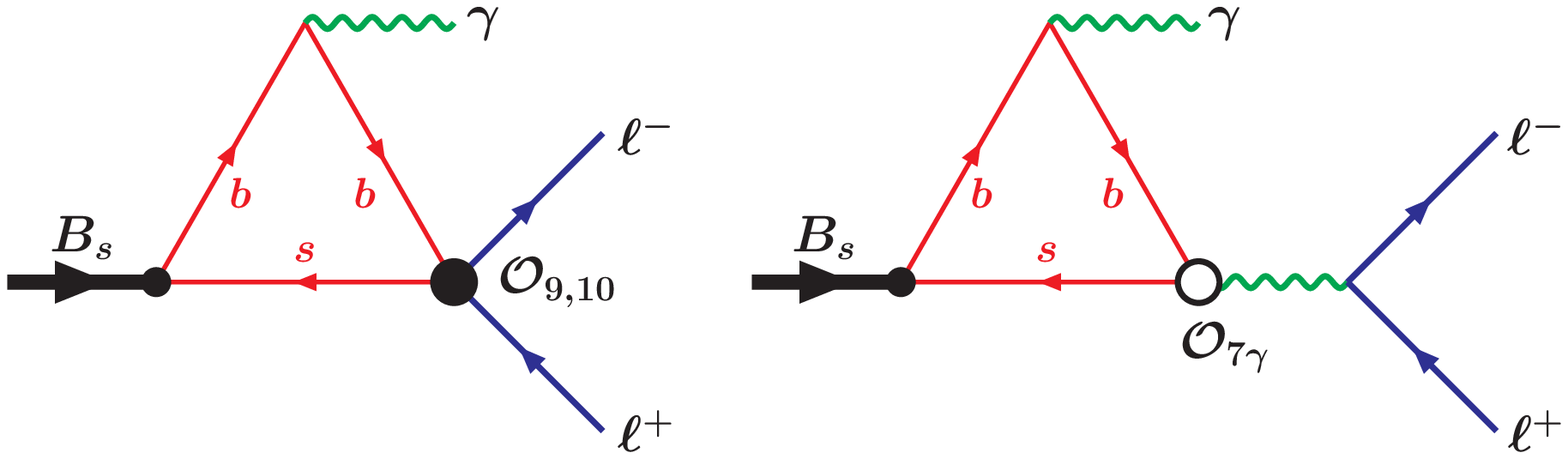} \\
\includegraphics[scale=0.5]{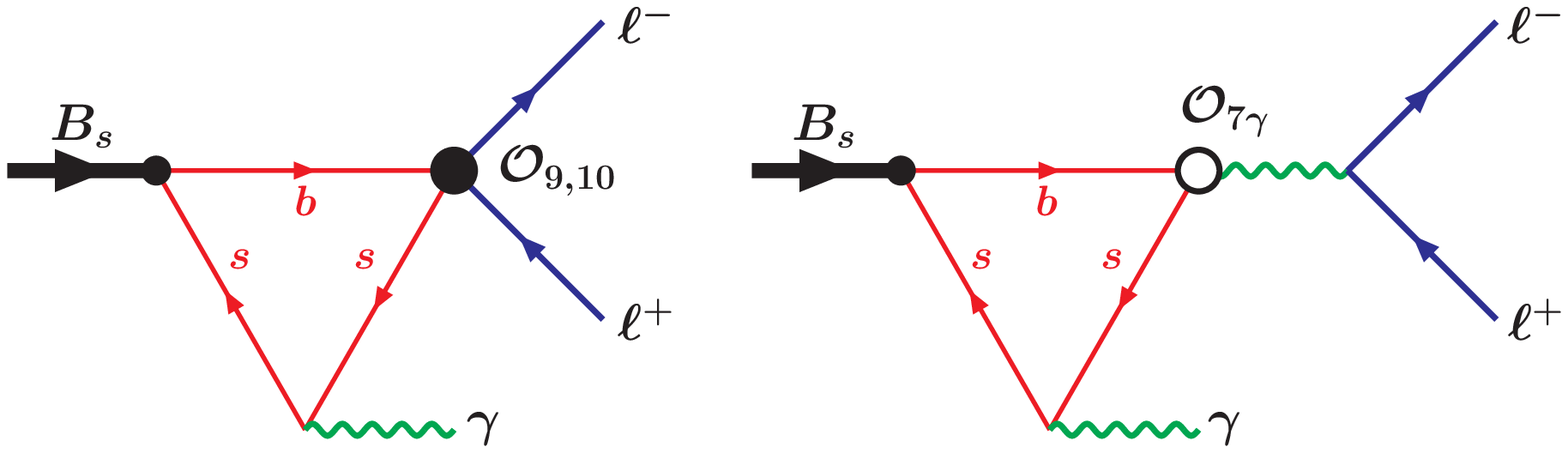} \\
\includegraphics[scale=0.5]{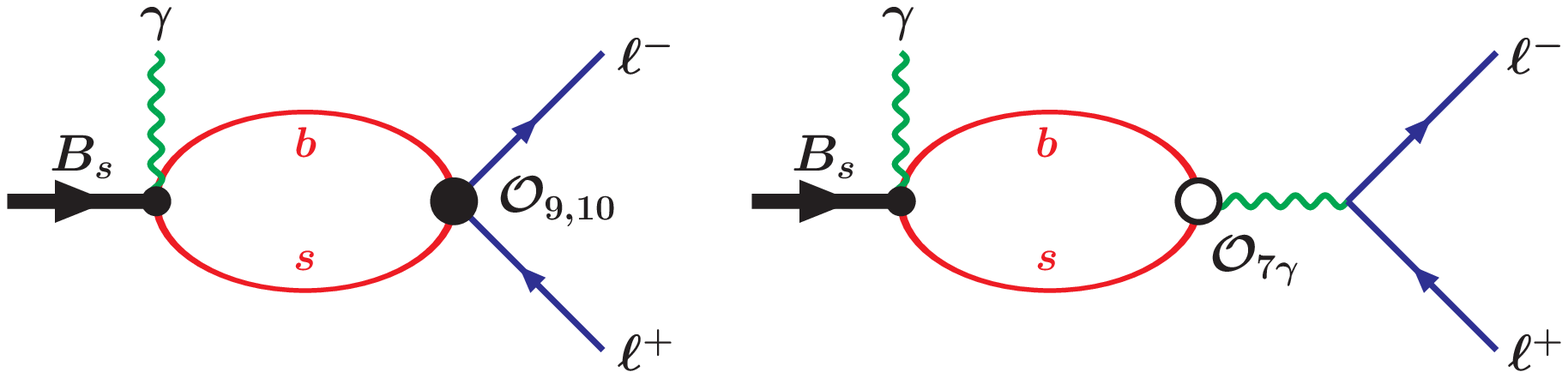} \\
\end{tabular}
\caption{Diagrams which contribute to the decays  $B_s\to \ell^+\ell^-\gamma$
with the real photon emitted from the quarks or the meson-quark vertex.}
\label{fig:D-1}
\end{figure*}

\begin{figure*}[htbp]
\centering
\begin{tabular}{c}
\includegraphics[scale=0.5]{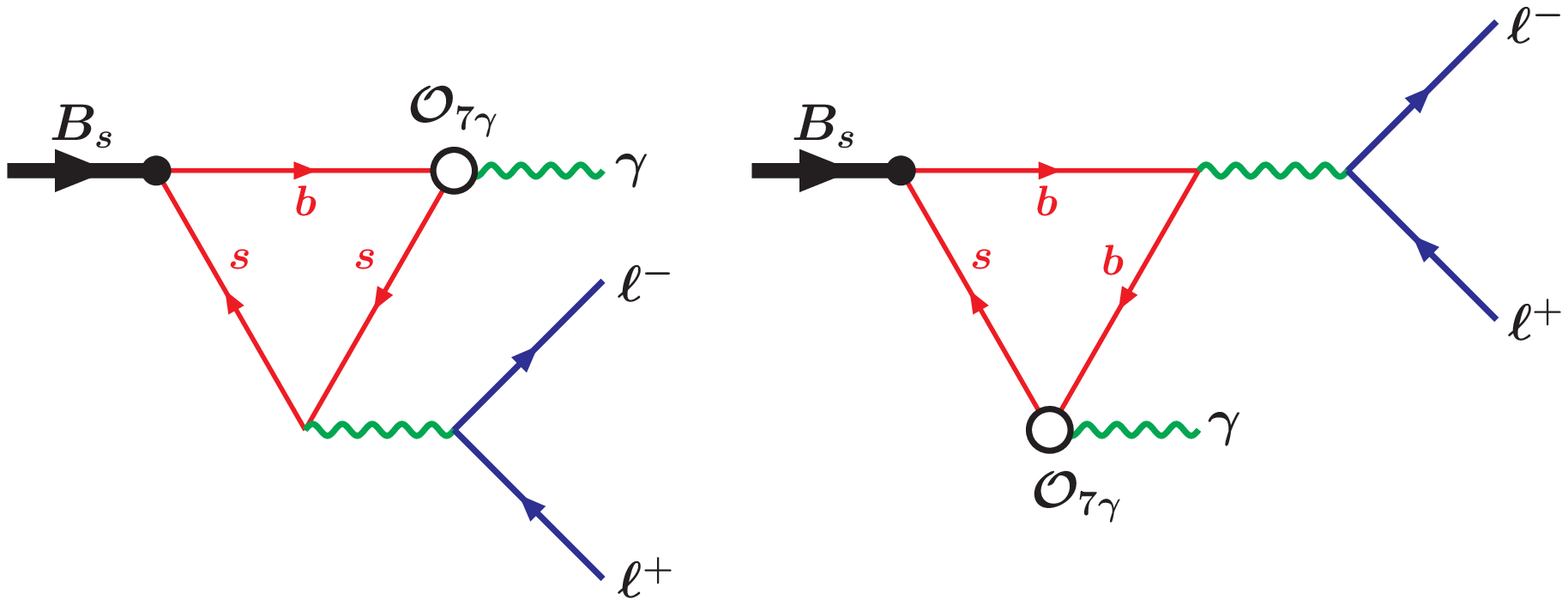} \\
\end{tabular}
\caption{Diagrams which contribute to the decays  $B_s\to \ell^+\ell^-\gamma$
with the real photon emitted from the penguin.}
\label{fig:D-2}
\end{figure*}

\begin{figure*}[htbp]
\centering
\begin{tabular}{c}
\includegraphics[scale=0.5]{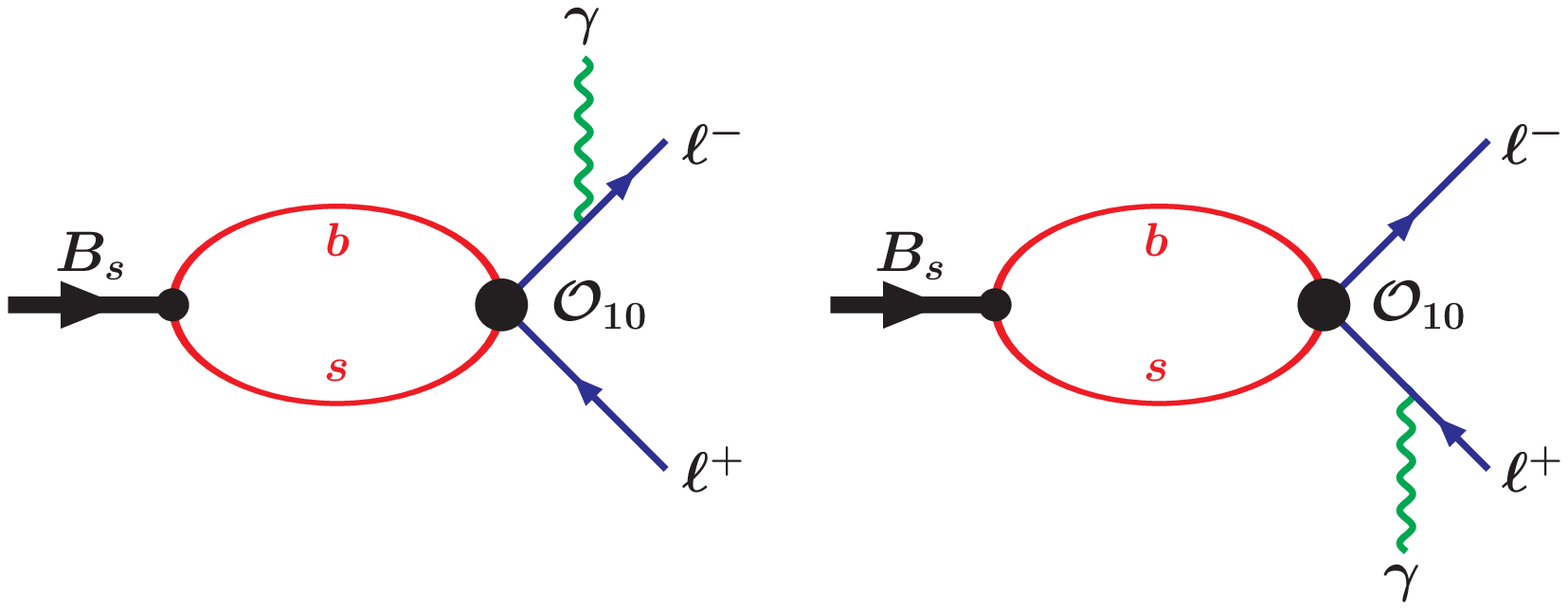} \\
\end{tabular}
\caption{Bremsstrahlung diagrams.} 
\label{fig:D-3}
\end{figure*}

The decays $B_s\to \ell^+\ell^-\gamma$ are described by three sets of diagrams
shown in Figs.~\ref{fig:D-1}--\ref{fig:D-3}. Diagrams from the first
set (Fig.~\ref{fig:D-1})  correspond to the case when the real photon is
emitted from the quarks or the meson-quark vertex.
The effective Hamiltonian describing the $b\to s\ell^+\ell^-$ weak transition 
is written as
\bea
{\mathcal H}_{\rm eff}^{b\to s\ell^+\ell^-} 
&=& \frac{G_F}{\sqrt{2}} \frac{\alpha_{\rm em}}{2\pi} \lambda_t    
\Big[
C_9^{\rm eff}\left(\bar{s}\gamma^\mu (1-\gamma_5) b\right) 
\left( \bar \ell \gamma_\mu \ell \right)
\nn
&-& \frac{2\tilde m_b}{q^2}C_7^{\rm eff} 
\left( \bar{s} i \sigma^{\mu \nu} q_\nu  (1+\gamma_5) b \right)  
\left( \bar \ell \gamma_\mu \ell \right)
\nn
&+& C_{10} \left(\bar{s}\gamma^\mu (1-\gamma_5) b\right)
\left(\bar \ell \gamma_\mu \gamma_5 \ell\right)
\Big],
\label{eq:eff-Ham-1}
\ena
where $\lambda_t = V_{tb} V^\ast_{ts}$, and $\tilde m_b$ is the QCD quark mass
which is different from the constituent quark mass $m_b$ used in our model. Here and in the following we denote the QCD quark masses with a tilde
to distinguish them from the constituent quark masses used in the model
[see Eq.~(\ref{eq:modelparam})]. 
The  Wilson coefficients $C_7^{\rm eff}= C_7 -C_5/3 -C_6$ and $C_{10}$
depend on the scale parameter $\mu$. The Wilson coefficient 
$ C_9^{\rm eff}$ effectively takes into account, first, the contributions 
from the four-quark operators $\ord_i$ ($i=1,...,6$) and, second, 
nonperturbative effects coming from the $c\bar c$-resonance contributions
which are as usual parametrized by the Breit-Wigner ansatz~\cite{Ali:1991is}:
\bea
C_9^{\rm eff} & = & C_9 + 
C_0 \left\{
h(\hat m_c,  s)+ \frac{3 \pi}{\alpha^2}  \kappa
         \sum\limits_{V_i = \psi(1s),\psi(2s)}
      \frac{\Gamma(V_i \rightarrow \ell^+ \ell^-) m_{V_i}}
{  {m_{V_i}}^2 - q^2  - i m_{V_i} \Gamma_{V_i}}
\right\} 
\nn
&&- \frac{1}{2} h(1,  s) \left( 4 C_3 + 4 C_4 +3 C_5 + C_6\right)  
\nn
&&- \frac{1}{2} h(0,  s) \left( C_3 + 3 C_4 \right) +
\frac{2}{9} \left( 3 C_3 + C_4 + 3 C_5 + C_6 \right),
\label{eq:C9eff}
\ena
where $C_0\equiv 3 C_1 + C_2 + 3 C_3 + C_4+ 3 C_5 + C_6$, 
$\hat m_c=\tilde m_c/M_{B_s}$, $s=q^2/M_{B_s}^2$, and $\kappa=1/C_0$.
Here,
\bea 
h(\hat m_c,  s) & = & - \frac{8}{9}\ln\frac{\tilde m_b}{\mu} 
- \frac{8}{9}\ln\hat m_c +
\frac{8}{27} + \frac{4}{9} x 
\nn
&& -\frac{2}{9} (2+x) |1-x|^{1/2} \left\{
\begin{array}{ll}
\left( \ln\left| \frac{\sqrt{1-x} + 1}{\sqrt{1-x} - 1}\right| - i\pi 
\right), &
\mbox{for } x \equiv \frac{4 \hat m_c^2}{ s} < 1, \nonumber \\
 & \\
2 \arctan \frac{1}{\sqrt{x-1}}, & \mbox{for } x \equiv \frac
{4 \hat m_c^2}{ s} > 1,
\end{array}
\right. 
\nn
h(0,  s) & = & \frac{8}{27} -\frac{8}{9} \ln\frac{\tilde m_b}{\mu} - 
\frac{4}{9} \ln s + \frac{4}{9} i\pi.
\nonumber 
\ena

The SM Wilson coefficients are taken from Ref.~\cite{Descotes-Genon:2013vna}.
They were computed at the matching scale $\mu_0=2 M_W$ and run down to 
the hadronic scale $\mu_b= 4.8$~GeV. Their numerical values are given
in Table~\ref{tab:input}. 
\begin{table}[htbp]
  \caption{NNLO Wilson coefficients at the scale $\mu_b=4.8$~GeV
    obtained in  Ref.~\cite{Descotes-Genon:2013vna}.}
\centering
\vskip 2mm
 \begin{tabular}{ccccccccc}
\hline
\quad $C_1$\quad  & \quad  $C_2$\quad  &
\quad $C_3$\quad  & \quad $C_4$\quad  &
\quad $C_5$\quad  & \quad  $C_6$\quad &
\quad $C^{\rm eff}_7$ \quad & \quad   $C_9$\quad  &
\quad $C_{10}$\quad  \\
\hline
\quad $-0.2632$ \quad&\quad $1.0111$\quad &
\quad $-0.0055$\quad  &\quad  $-0.0806$\quad  &
\quad 0.0004\quad  &\quad  0.0009\quad  &
\quad $-0.2923$\quad  &
\quad   4.0749\quad  &\quad  $-4.3085$\quad  \\
\hline
 \end{tabular}
\label{tab:input}
\end{table}

We use the bare $c$-quark mass corresponding to the “running” mass
$\tilde m_c = \bar m_c (\mu~=\bar m_c)=1.27\pm 0.03$~GeV in
the $\overline{MS}$ scheme (for a review, see ``Quark masses'' in PDG
\cite{Tanabashi:2018oca}).
Note that the $\tilde m_c$ appears only in the charm-loop function
$h(\hat m_c,s)$ via the logarithm. Therefore, uncertainties related to
the choice of the scale parameter $\mu$ are small. 
For the bare $b$-quark mass we use the central value of
$ \tilde m_b =m^{1S}_b = 4.68\pm 0.03$~GeV
obtained in the $1S$ mass scheme; see Ref.~\cite{Bauer:2004ve}.
This value is close to the pole b-mass which was used in
the Wilson coefficients $C_i(\mu_b).$
Finally, the values of $\alpha_{\rm em}(M_Z)=1/128.94$ and
$\lambda_t=|V_{tb} V^\ast_{ts}| = 0.040$ are taken from PDG
\cite{Tanabashi:2018oca}.

Diagrams from the second set (Fig.~\ref{fig:D-2})  represent the case when the real photon is emitted from the magnetic penguin operator.
The effective Hamiltonian describing the $b\to s\gamma$ electroweak transition 
is written as
\be
{\mathcal H}_{\rm eff}^{b\to s\gamma}  = 
-\frac{G_F}{\sqrt{2}} \lambda_t   
C_7^{\rm eff}\frac{e \tilde m_b}{8\pi^2}
\left( \bar{s} \sigma_{\mu \nu} (1+\gamma_5) b \right) F^{\mu\nu}.  
\en

Diagrams from the first two sets contribute to the structure-dependent (SD) part of the decay amplitude. They can be parametrized by a set of invariant form factors. In order to define the form factors, we specify our choice for the momenta in the decays as follows:
\be
B_s(p_1)\to \gamma(p_2)+\ell^+(k_+) + \ell^-(k_-),
\en
where $p_1=p_2+k_++k_-$ and $p_1-p_2=k_++k_-\equiv q$,  with $p_1^2=M^2_{B_s}$, 
$p_2^2=0$, $\epsilon^\dagger_2\cdot p_2=0$, and   $k_+^2=k_-^2=m_\ell^2$. 

We will use the definition of the $B_s\to\gamma$ transition form factors
given, for instance, in Ref.~\cite{Melikhov:2004mk}:
\bea
\la \gamma(p_2,\epsilon_2) |\bar s \gamma^\mu  b | B_s(p_1) \ra
&=&
e \epsilon_{2\alpha}^{\dagger} \varepsilon^{\mu\alpha p_1p_2} 
   \frac{F_V(q^2)}{M_{B_s}},
\nn
\la \gamma(p_2,\epsilon_2) |\bar s \gamma^\mu\gamma_5  b| B_s(p_1) \ra
&=&
i e \epsilon_{2\alpha}^{\dagger} ( g^{\mu\alpha} p_1p_2 - p_1^\alpha p_2^\mu )
   \frac{F_A(q^2)}{M_{B_s}},
\nn
\la \gamma(p_2,\epsilon_2) |\bar s \sigma^{\mu q}  b| B_s(p_1) \ra
&=&
i e \epsilon_{2\alpha}^{\dagger} \varepsilon^{\mu\alpha p_1p_2} F_{TV}(q^2),
\nn
\la \gamma(p_2,\epsilon_2) |\bar s \sigma^{\mu q}\gamma_5  b| B_s(p_1) \ra
&=&
e \epsilon_{2\alpha}^{\dagger} ( g^{\mu\alpha} p_1p_2 - p_1^\alpha p_2^\mu )F_{TA}(q^2).
\label{eq:ffdef}
\ena
Here we use the short notations $\sigma^{\mu q} \equiv \sigma^{\mu \beta} q_{\beta}$ and
$ \varepsilon^{\mu\alpha p_1p_2} \equiv \varepsilon^{\mu\alpha\beta\delta} p_{1 \beta}p_{2 \delta}$. 

The form factors gain contributions from the diagrams in Figs.~\ref{fig:D-1} and~\ref{fig:D-2} in the following manner: 
\bea
F_V &=& M_{B_s} \left[ e_b \tilde F_V^{b\gamma b} + e_s \tilde F_V^{s\gamma s} \right],
\nn
F_A &=& M_{B_s} \left[  e_b \tilde F_A^{b\gamma b} + e_s \tilde F_A^{s\gamma s} 
                  + e_b \tilde F_A^{{\rm bubble}-b} + e_s \tilde F_A^{{\rm bubble}-s}  \right],
\nn
F_{TV} &=&   e_b \tilde F_{TV}^{b\gamma b} + e_s \tilde F_{TV}^{s\gamma s} 
                  + e_b \tilde F_{TV}^{b(\bar\ell\ell)b} + e_s \tilde F_{TV}^{s(\bar\ell\ell)s} ,
\nn
F_{TA} &=&   e_b \tilde F_{TA}^{b\gamma b} + e_s \tilde F_{TA}^{s\gamma s}
          + e_b \tilde F_{TA}^{{\rm bubble}-b} + e_s \tilde F_{TA}^{{\rm bubble}-s} 
                  + e_b \tilde F_{TA}^{b(\bar\ell\ell)b} + e_s \tilde F_{TA}^{s(\bar\ell\ell)s}.
\label{eq:ffcont}                  
\ena

The process with the virtual photon emitted from the light $s$ quark is described by the diagram in Fig.~\ref{fig:D-2} (left panel). The physical region for  $q^2$
in the decays $B_s\to \gamma \ell^+\ell^-$ is extended up to $q^2_{\rm max} = M^2_{B_s}$, which is much higher than the branch-point value $q^2=4m_s^2$.
In this case, the form factor $\tilde F_{TV/TA}^{s(\bar\ell\ell)s}$ cannot be directly calculated in our model due to the appearance of hadronic singularities associated with the light vector meson resonances. In order to describe this amplitude, we follow the authors of Ref.~\cite{Kozachuk:2017mdk} in using the gauge-invariant version~\cite{Melikhov:2003hs} of the vector meson dominance~\cite{Sakurai:1960ju, GellMann:1961tg, Gounaris:1968mw}, 
\be
\tilde F_{TV/TA}^{s(\bar\ell\ell)s}(q^2)
=\tilde F_{TV/TA}^{s(\bar\ell\ell)s}(0)-\sum\limits_{V} 2f_V^{\rm e.m.} G^T_1(0)\frac{q^2/M_V}{q^2-M_V^2+iM_V\Gamma_V},
\label{eq:FF-BW}
\en
where $\Gamma_V$ and $M_V$ are the decay width and mass of the vector meson resonance, and $G^T_1(0)$ is one of the tensor form factors for the $B_s \to V$ transition, defined as follows~\cite{Ivanov:2016qtw, Ivanov:2017hun,Tran:2018kuv}:
\be
\langle V(p_2,\epsilon_2)|\bar{s}\sigma^{\mu\nu}b|B_s(p_1)\rangle
=\epsilon^\dagger_{2\alpha}\Big[
\varepsilon^{P\mu\nu\alpha}G_1^T(q^2)
+ \varepsilon^{q\mu\nu\alpha}G_2^T(q^2)
+\varepsilon^{Pq\mu\nu}P^\alpha\frac{G_0^T(q^2)}{(M_{B_s}+M_V)^2}
\Big].
\en
All parameters necessary for the form factor definition in Eq.~(\ref{eq:FF-BW}) are calculated in our model and are given by
\be
\begin{tabular}{ l c c }
$\tilde F_{TV/TA}^{s(\bar\ell\ell)s}(0)$ \quad & \quad $f_\phi$ (GeV) \quad & \quad $G_1^T(0)$ \quad 
\\\hline
 \quad 0.120 \quad & \quad 0.227 \quad & \quad 0.266 \quad 
\end{tabular}
.
\label{eq:VMD-para}
\en
Note that the electromagnetic decay constant  is related to the leptonic decay constant by the relation $f_\phi^{\rm e.m.}=-\frac13 f_\phi$. Regarding the light resonances, here we consider only the main contribution from the ground-state $\phi$ meson. 
It is interesting to note that our result for $G_1^T(0)$ is equal to the value $0.27\pm 0.01$ obtained by the authors of Ref.~\cite{Kozachuk:2017mdk}.

Finally, the SD part of the amplitude is written in terms of the form factors as follows: 
\bea
\mathcal{M}_{\rm SD} &=& \frac{G_F}{\sqrt{2}}\frac{\alpha_{\rm em} \lambda_t}{2\pi} e\epsilon^\ast_{2 \alpha}
\Big\{
\Big[ \varepsilon^{\mu\alpha p_1p_2} \frac{F_V(q^2)}{M_{B_s}} - i T_1^{\mu\alpha}\frac{F_A(q^2)}{M_{B_s}}\Big] 
\left( C_9^{\rm eff}\bar\ell\gamma_\mu\ell +  C_{10}\bar\ell\gamma_\mu\gamma_5\ell \right)
\nn
&&
\phantom{ \frac{G_F}{\sqrt{2}}\frac{\alpha_{\rm em} \lambda_t}{2\pi} e\epsilon_{2 \alpha}}
+ \Big[ \varepsilon^{\mu\alpha p_1p_2} F_{TV}(q^2) - i T_1^{\mu\alpha} F_{TA}(q^2)\Big]
  \frac{2\tilde m_b}{q^2} C_{7}^{\rm eff}  \bar\ell\gamma_\mu\ell
\Big\},
\ena
where $T_1^{\mu\alpha} \equiv ( g^{\mu\alpha} p_1p_2 - p_1^\alpha p_2^\mu )$.

The structure-independent part of the amplitude (bremsstrahlung) is described by the diagrams in Fig.~\ref{fig:D-3}. 
Only the operator $\ord_{10}$ contributes
to this process, and it effectively gives the leptonic decay constant $f_{Bs}$. One has
\be
\mathcal{M}_{\rm BR} = -i \frac{G_F}{\sqrt{2}}\frac{\alpha_{\rm em} \lambda_t}{2\pi} e\epsilon^\ast_{2 \alpha}
(2m_\ell f_{Bs} C_{10})
\bar u({\boldmath{k_-}})
\Big[  \frac{\gamma^\alpha \not\! p_1}{ t -m^2_\ell} 
     - \frac{\not\! p_1 \gamma^\alpha}{u -m^2_\ell}\Big] \gamma_5
v({\boldmath{k_+}}).
\label{eq:BR}
\en
Here, $t=(p_2+k_-)^2=(p_1-k_+)^2$, $u=(p_2+k_+)^2=(p_1-k_-)^2$, and $s=q^2$ so that
$s+t+u=M_{B_s}^2+2m^2_\ell$. The variable $t$ varies in the interval
$t_-\le t \le t_+$, where the bounds $t_{\pm}$ are given by 
\be
t_{\pm} = m^2_\ell +\frac12 (M_{B_s}^2-s) [1 \pm \beta(s)],
\qquad \beta(s) = \sqrt{1-\frac{4 m^2_\ell}{s}}.
\en
One can see that  $t_{\pm}=m^2_\ell$ at minimum recoil $s=q^2=M_{B_s}^2$,
which leads to the infrared pole in Eq.~(\ref{eq:BR}).
A cut in the photon energy is required. In the $B_s$ center-of-mass system
one has
\be 
E_\gamma = \frac{M_{B_s}}{2}\left(1 - \frac{q^2}{M_{B_s}^2}\right) \ge E_{\gamma\,\rm min}.
\label{eq:Egam}
\en

Note that there are also weak annihilation diagrams with a $u(c)$ anomalous triangle
in addition to the above diagrams. However, the contribution from these
diagrams is much smaller than that from other diagrams as was shown in
Ref.~\cite{Melikhov:2004mk}. Therefore, we will drop these types of diagrams in what follows.

A few remarks should be made with respect to the calculation of the Feynman diagrams
in our approach. The SD part of the matrix element is described
by the diagrams in Figs.~\ref{fig:D-1} and \ref{fig:D-2}. These diagrams do not include
ultraviolet divergences because the hadron-quark vertex functions drop off exponentially
in the Euclidean region. The loop integration is performed by using the Fock-Schwinger
representation for the quark propagators, and the exponential form for the meson-quark vertex 
functions. The tensorial integrals are calculated by using the differential technique.
The final expression for the SD part is represented as a sum of 
products of Lorentz structures and the corresponding invariant form factors.
The form factors are described by threefold  integrals in such a way that one
integration is over a dimensional parameter $t$ (proper time), which proceeds from zero to infinity, 
and two others are over dimensionless Schwinger parameters. The possible branch points and cuts
are regularized by introducing the cutoff at the upper limit of the integration over proper time. 
The final integrals are calculated numerically by using the \textsc{fortran} codes.

Then one can check that the final expression for the SD part of the amplitude
is gauge invariant. Technically, it means that in addition to the gauge-invariant
structures $\varepsilon^{\mu\alpha p_1p_2}$ and   
$T_1^{\mu\alpha} = ( g^{\mu\alpha} p_1p_2 - p_1^\alpha p_2^\mu )$, the amplitude has also the
non-gauge-invariant pieces $g^{\mu\alpha}$ and $p_1^\mu p_1^\alpha$. We have checked numerically that the form factors corresponding to the non-gauge-invariant part
 vanish for arbitrary momentum transfer squared $q^2$.
 
\section{Differential decay rate}
\label{sec:dis}

The twofold decay distribution is written as
\be
\frac{d\Gamma}{ds dt} = \frac{1}{2^8 \pi^3 M_{B_s}^3}\sum_{\rm pol} |\mathcal{M}|^2, \qquad
\mathcal{M} = \mathcal{M}_{\rm SD} + \mathcal{M}_{\rm BR},
\label{eq:width-2}
\en
where $\sum_{\rm pol}$ denotes the summation over polarizations of both the photon and leptons.
The physical region was discussed in the previous section, which reads $4 m^2_\ell\le s\equiv q^2 \le M_{B_s}^2$, and 
$t_-\le t \le t_+$. It is more convenient to write the final result for the twofold decay distribution
in terms of dimensionless momenta and masses:
\be
\hat{X} \equiv \frac{X}{M^2_{B_s}}\quad (X = s,t,u), \quad 
\hat{Y} \equiv \frac{Y}{M_{B_s}}\quad (Y=m_\ell, \tilde{m}_q, f_{B_s}), \quad\text{etc.}
\en

The decay distribution is written as a sum of the SD part, the bremsstrahlung (BR), and the interference (IN) ones as follows:
\be
\frac{d\Gamma}{d\Hs d \Ht} = N_t
\Big(  \frac{d\Gamma_{\rm SD}}{d\Hs d \Ht}  +   \frac{d\Gamma_{\rm BR}}{d\Hs d \Ht} +  \frac{d\Gamma_{\rm IN}}{d\Hs d \Ht} \Big),
\qquad N_t \equiv \frac{G_F^2\alpha_{\rm em}^3 M_{B_s}^5 |\lambda_t|^2}{2^{10}\pi^4}.
\label{eq:two-fold}
\en
One has
\bea
 \frac{d\Gamma_{\rm SD}}{d\Hs d \Ht} &=& \Hx^2 B_0 + \Hx(\Hu-\Ht)B_1 + (\Hu-\Ht)^2 B_2,
\\
&& B_0 = (\Hs + 4 \Hml^2)\Delta F-8 \Hml^2 C_{10}^2 ( F_V^2 + F_A^2), 
\nn
&& B_1 = 8\big[  \Hs C_{10}\re(C^{\rm eff}_9) F_V F_A 
              + \Hmb C_{7}^{\rm eff}C_{10}( F_V \re(F_{TA}) + F_A \re(F_{TV})) \big],  
\nn
&& B_2 = \Hs \Delta F,
\nn
&& \Delta F = (|C^{\rm eff}_9|^2 + C_{10}^2) (F_V^2 +F_A^2)
 +\Big(\frac{2 \Hmb}{\Hs}\Big)^2 (C_{7}^{\rm eff})^2 (|F_{TV}|^2 +|F_{TA}|^2)
\nn
&& \phantom{\Delta F =}
 +\Big(\frac{4 \Hmb}{\Hs}\Big)C_{7}^{\rm eff} 
[F_V \re(C^{\rm eff}_9 F_{TV}) + F_A \re(C^{\rm eff}_9 F_{TA})],
\nn
 \frac{d\Gamma_{\rm BR}}{d\Hs d \Ht} &=& (8 \HfBs)^2 \Hml^2 C_{10}^2
\Big[ \frac12(1 + \Hs^2)\hat D_u \hat D_t - ( \Hx\Hml \hat D_u \hat D_t)^2 \Big],
\\
 \frac{d\Gamma_{\rm IN}}{d\Hs d \Ht} &=& -16 \HfBs \Hml^2 \Hx^2 \hat D_u \hat D_t\nn
&&\times\Big[ \frac{2 \Hx \Hmb}{\Hs}C_{10}C_{7}^{\rm eff}\re(F_{TV}) + \Hx C_{10}\re(C^{\rm eff}_9)F_V
+(\Hu-\Ht)C_{10}^2 F_A\Big], 
\ena
where $\Hx=1-\Hs$, $\hat D_t = 1/(\Ht - \Hml^2)$, and $\hat D_u = 1/(\Hu - \Hml^2)$.
We have checked that all the expressions above are in agreement with those given in 
Ref.~\cite{Kozachuk:2017mdk}.

After integrating out the variable $\hat{t}$ we obtain the following analytic expressions for the differential decay rate:
\bea
\frac{d\Gamma_{\rm SD}}{d\Hs} 
&=& 
N_t\Hx^3\beta\big[B_0 + \tfrac13 \beta^2 B_2 \big],
\label{eq:SD}\\
\frac{d\Gamma_{\rm BR}}{d\Hs} 
&=& 
N_t(8 \HfBs)^2\Hml^2 C^2_{10}
\Big[ 
 \frac{1+\Hs^2}{\Hx} \ln\Big(\frac{1+\beta}{1-\beta}\Big)
\nn
&& \phantom{N_t (8 \HfBs)^2\Hml^2 C^2_{10}\Big[}
 -\frac{8\Hml^2}{\Hx} 
  \Big( \frac{\beta}{1-\beta^2}
  +\tfrac12 \ln\Big(\frac{1+\beta}{1-\beta}\Big) \Big)
\Big],
\label{eq:BR-1}\\
\frac{d\Gamma_{\rm IN}}{d\Hs} 
&=& 
-N_t 32 \HfBs\Hml^2\Hx^2\ln\Big(\frac{1+\beta}{1-\beta}\Big)
\Big[ 
C_{10}\re (C^{\rm eff}_9)F_V + \frac{2\Hmb}{\Hs}C_{7}^{\rm eff}C_{10}\re(F_{TV})
\Big],
\label{eq:IN}
\ena
where $\beta =\sqrt{1-4\Hml^2/\Hs}$. 

\section{Numerical results}
\label{sec:result}

\begin{figure}[ht]
\begin{center}
\begin{tabular}{lr} 
\includegraphics[scale=0.5]{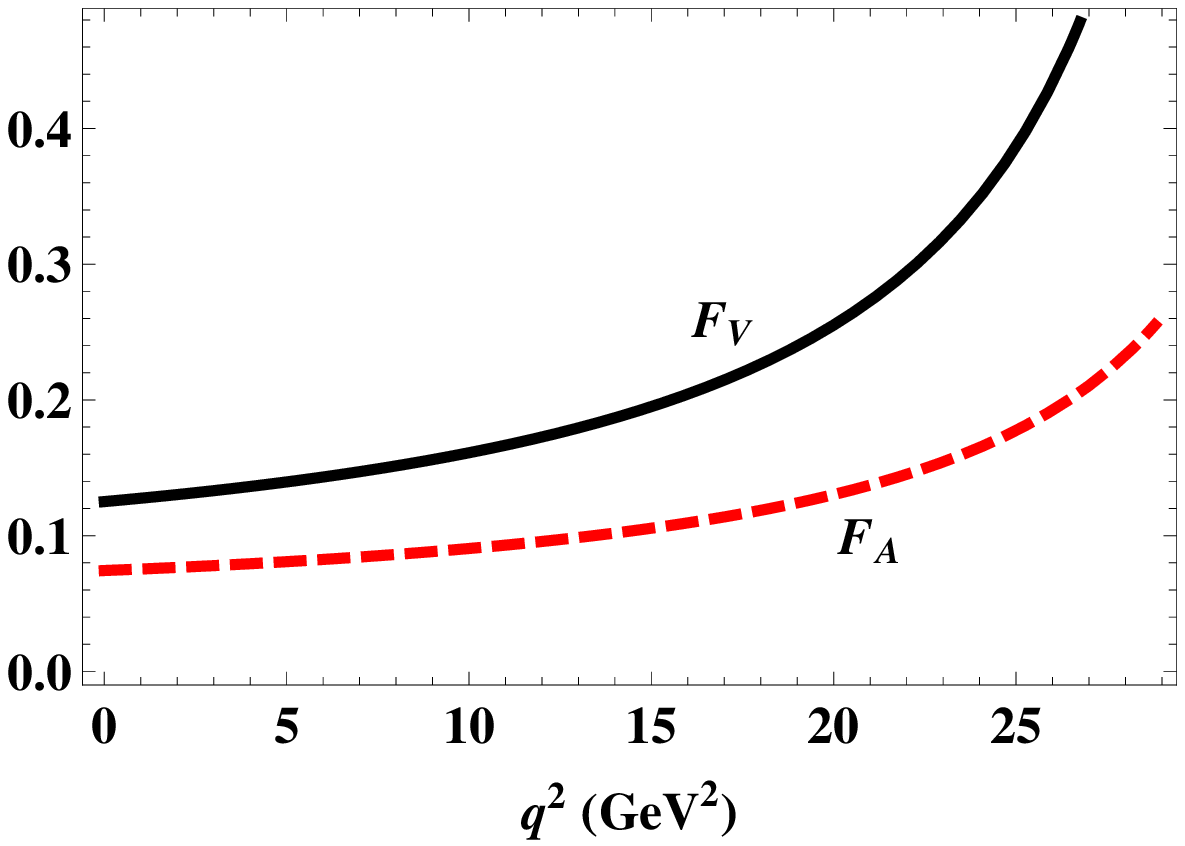} &
\includegraphics[scale=0.5]{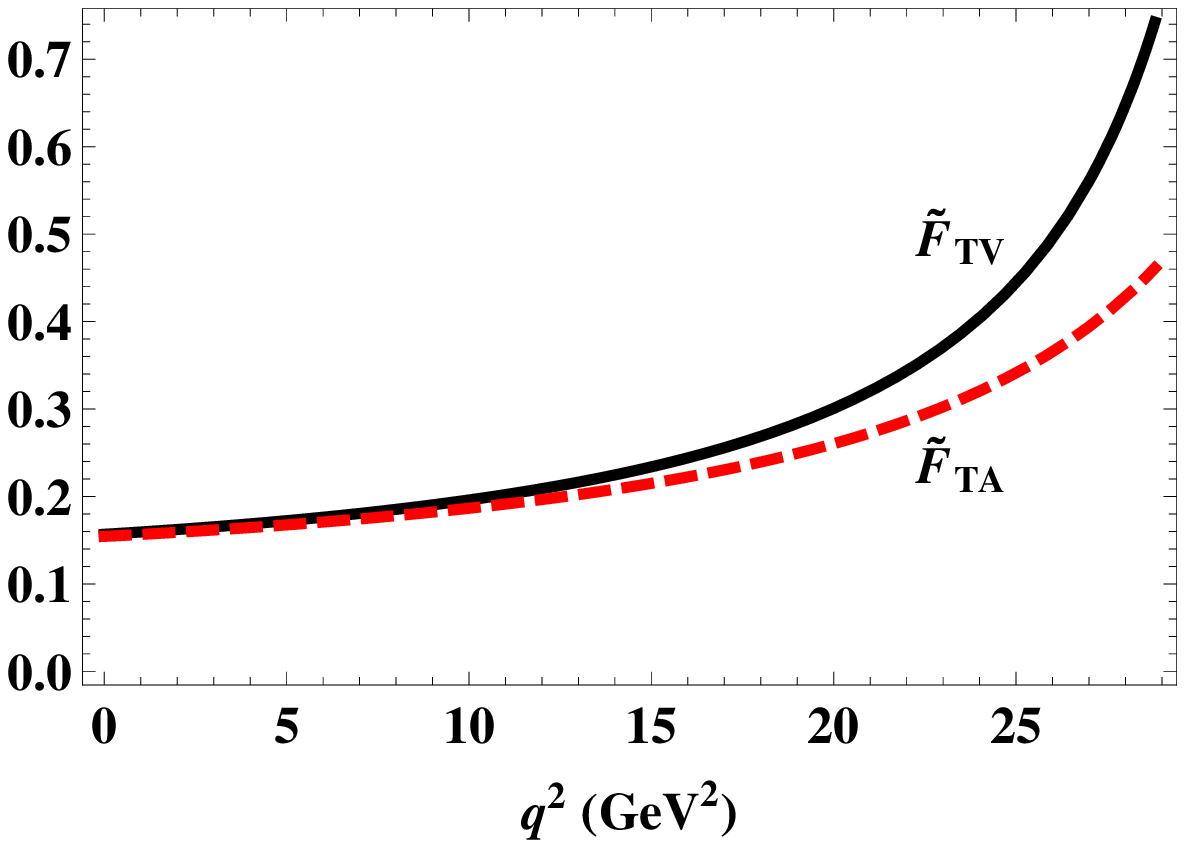}\\
\includegraphics[scale=0.5]{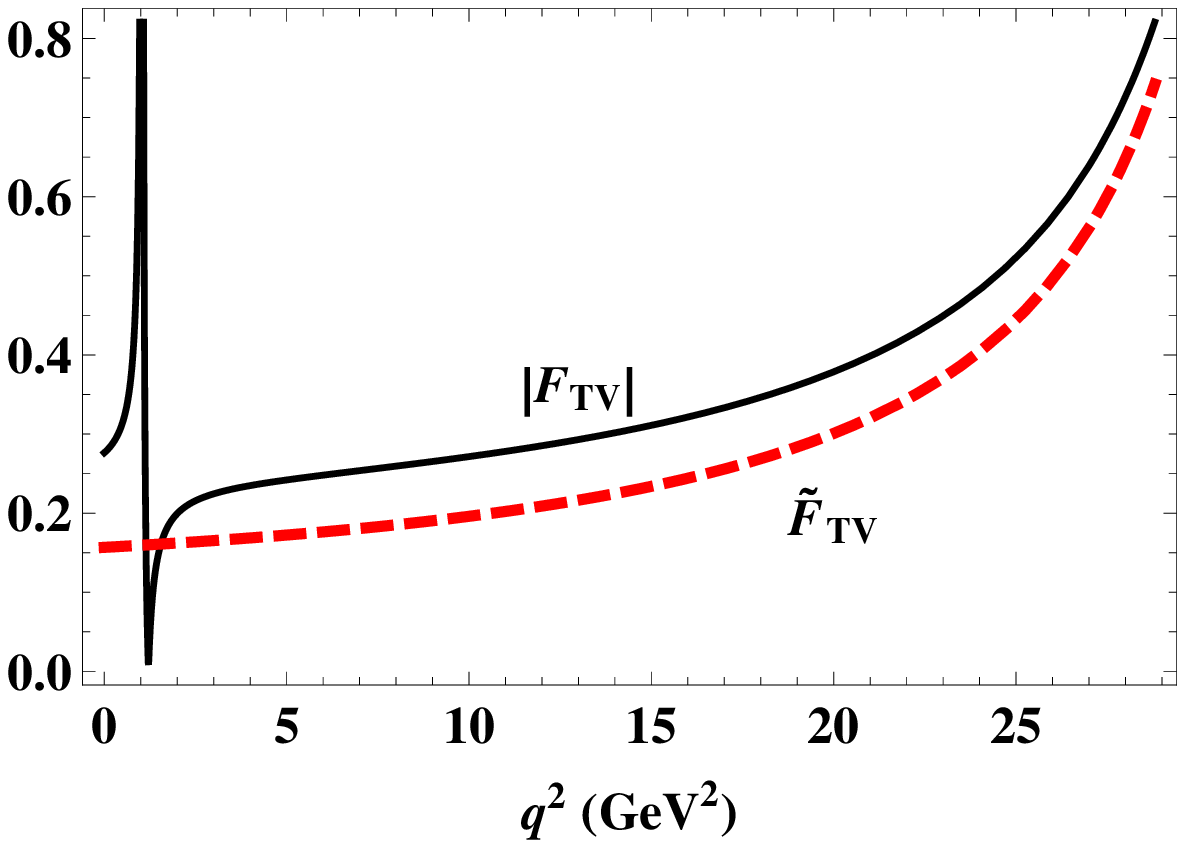} &
\includegraphics[scale=0.5]{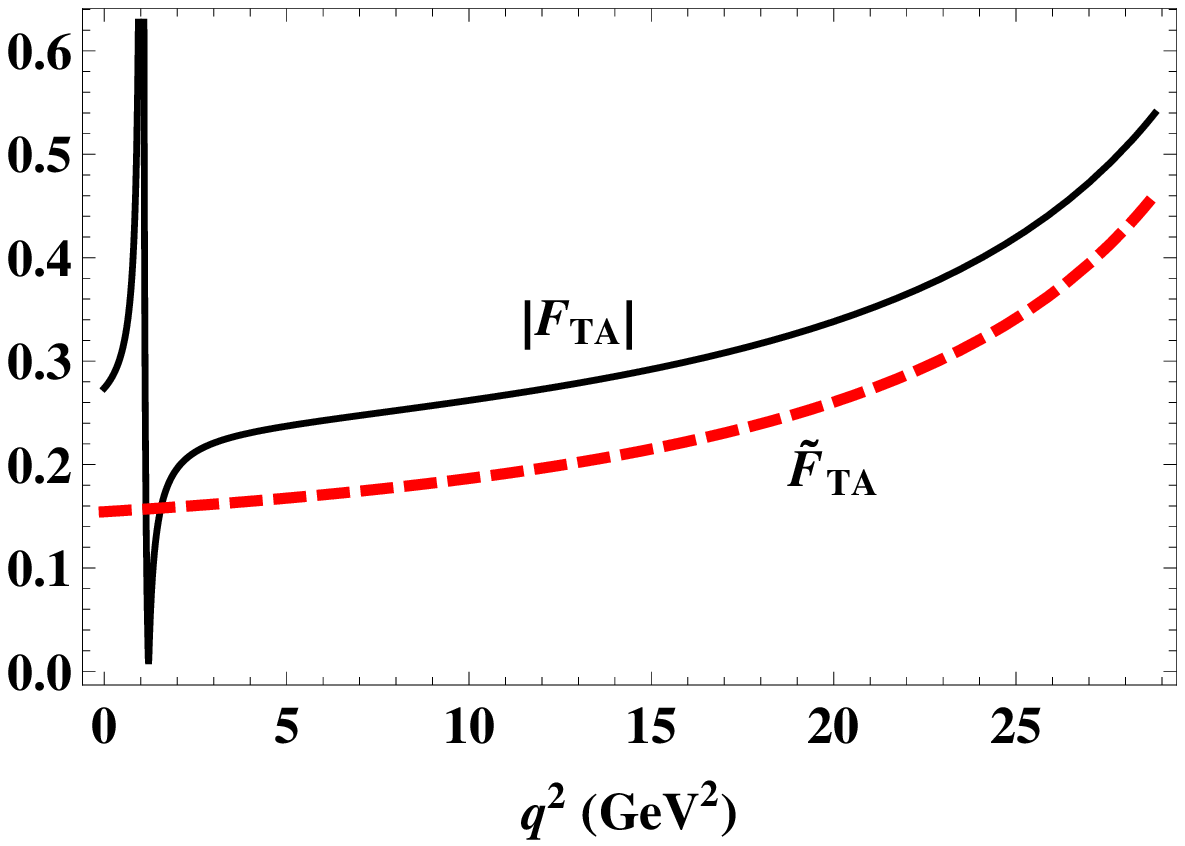}
\end{tabular}
\end{center}
\caption{\label{fig:ff}
Form factors for the $B_s \to \gamma$ transition (see text for more details).
}
\end{figure}

In Fig.~\ref{fig:ff} we present the $q^2$ dependence of the calculated form
factors with the fixed values of the model parameters
given in Eq.~(\ref{eq:modelparam}) in the full kinematical region
$0\le q^2\le M^2_{B_s}$. 
Here, the form factors $\tilde{F}_{TV}$ and $\tilde{F}_{TA}$ are defined as
follows:
\be
\tilde{F}_{TV/TA}(q^2)=F_{TV/TA}(q^2)-e_s \tilde{F}_{TV/TA}^{s(\bar{l}l)s}.
\en
One sees that $\tilde{F}_{TV}$ and $\tilde{F}_{TA}$ are real and are made of contributions from all diagrams except for the one with the virtual photon emitted from the $s$ quark. The total form factors $F_{TV}$ and $F_{TA}$ are complex due to the parametrization of 
$\tilde{F}_{TV/TA}^{s(\bar{l}l)s}$ in Eq.~(\ref{eq:FF-BW}). In Fig.~\ref{fig:ff} (lower panels) we also plot the absolute values $|F_{TV}|$ and $|F_{TA}|$ together with $\tilde{F}_{TV}$ and $\tilde{F}_{TA}$ for comparison.
%

The results of our numerical calculations for the form factors $F_V$, $F_A$,
$\tilde{F}_{TV}$, and $\tilde{F}_{TA}$ can be approximated 
with high accuracy by the double-pole parametrization
\be
F(q^2)=\frac{F(0)}{1-a\hat s+b\hat s^2}, \qquad 
\hat s=\frac{q^2}{M_{B^\ast_s}^2},
\label{eq:ff_approx}
\en
with the relative error less than 1$\%$.
The parameters $F(0)$, $a$, and $b$ are listed  in Table~\ref{tab:apprff}.
For completeness we also list here the values of the form factors at
zero recoil $(q^2_{\rm max})$.
\begin{table}[ht]
\caption{Parameters of the approximated $B_s\rightarrow \gamma$ form factors 
 and their values at zero recoil.} 
\begin{center}
\begin{tabular}{crrrr}
\hline
\hline
   &\qquad $F_V$ \qquad & \qquad $F_A$ \qquad & \qquad $\tilde{F}_{TV}$ \qquad & \qquad $\tilde{F}_{TA}$ \qquad \\
\hline
$F(0)$ 	&  0.13         &        0.074        &        0.16          &    0.15  
\\
$a$    	&  0.56         &        0.42         &        0.47          &    0.41
\\
$b$    	& $-0.27$       &        $-0.31$      &        $-0.34$       &   $-0.27$
\\
$F(q^2_{\rm max})$  &    0.67       &   0.26        &   0.74        &  0.46
\\
\hline
\hline
\end{tabular}
\label{tab:apprff} 
\end{center}
\end{table}

\begin{figure}[ht]
\begin{center}
\begin{tabular}{ll} 
\includegraphics[scale=0.5]{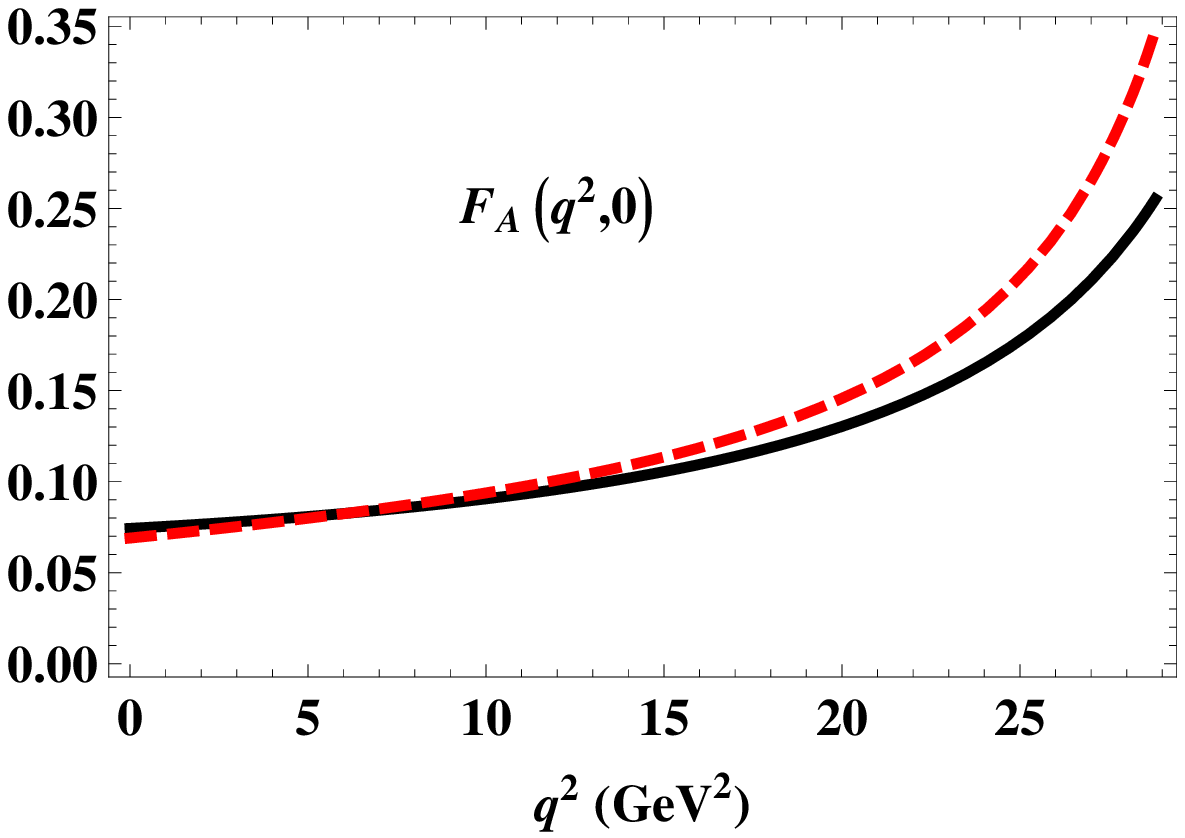} &
\includegraphics[scale=0.5]{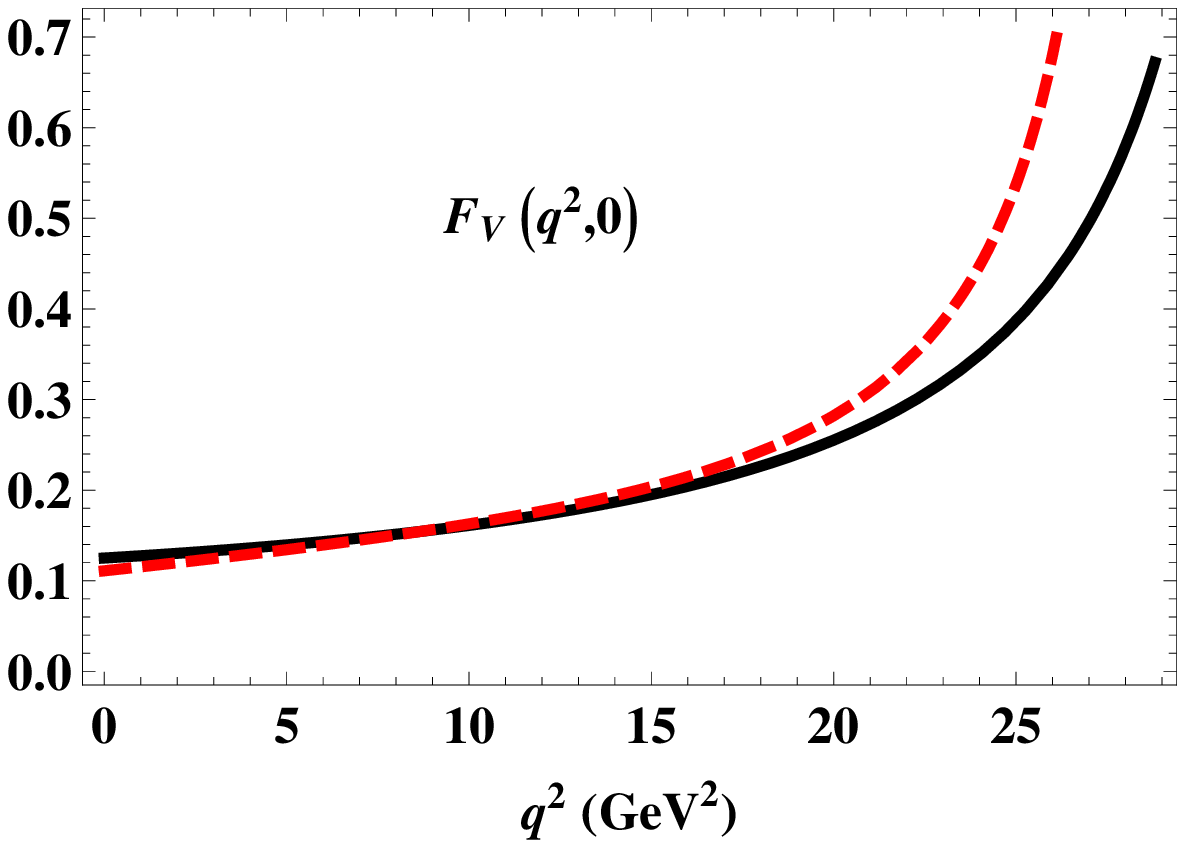} \\ 
\includegraphics[scale=0.5]{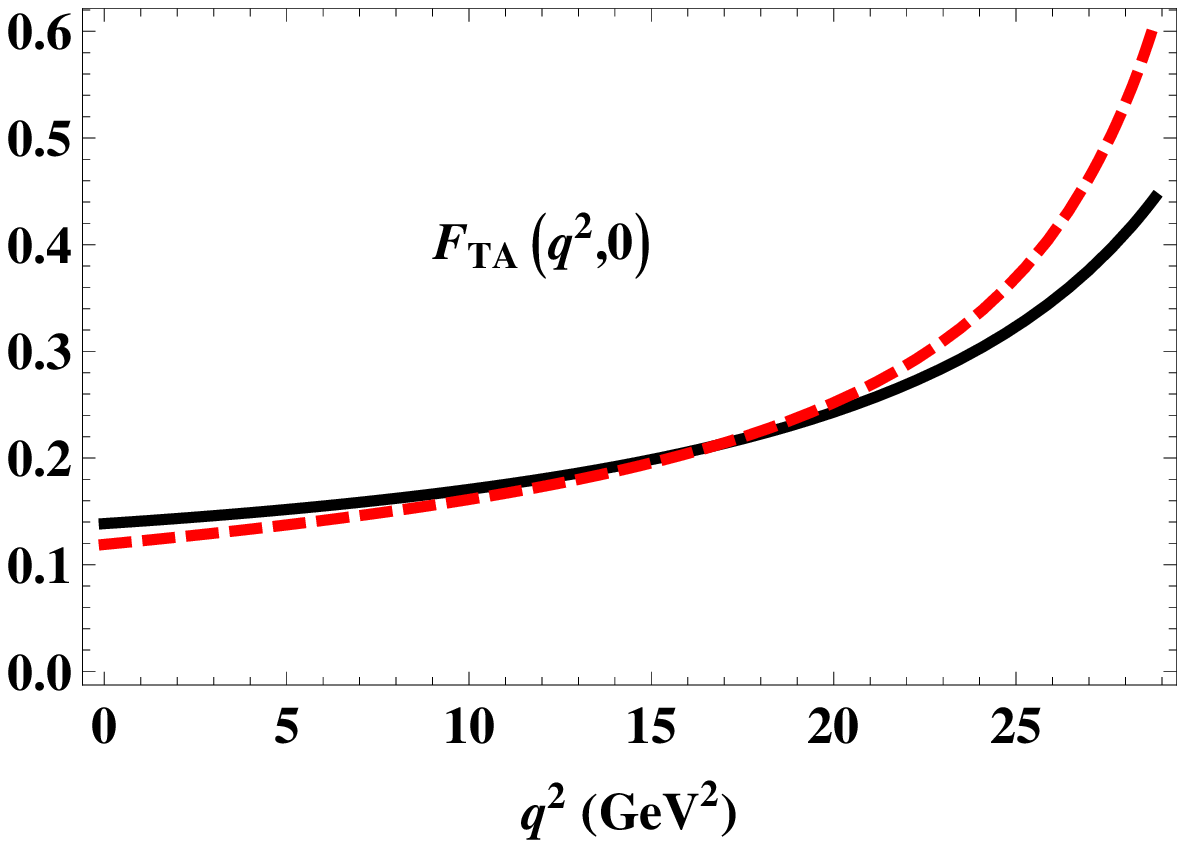} &
\includegraphics[scale=0.5]{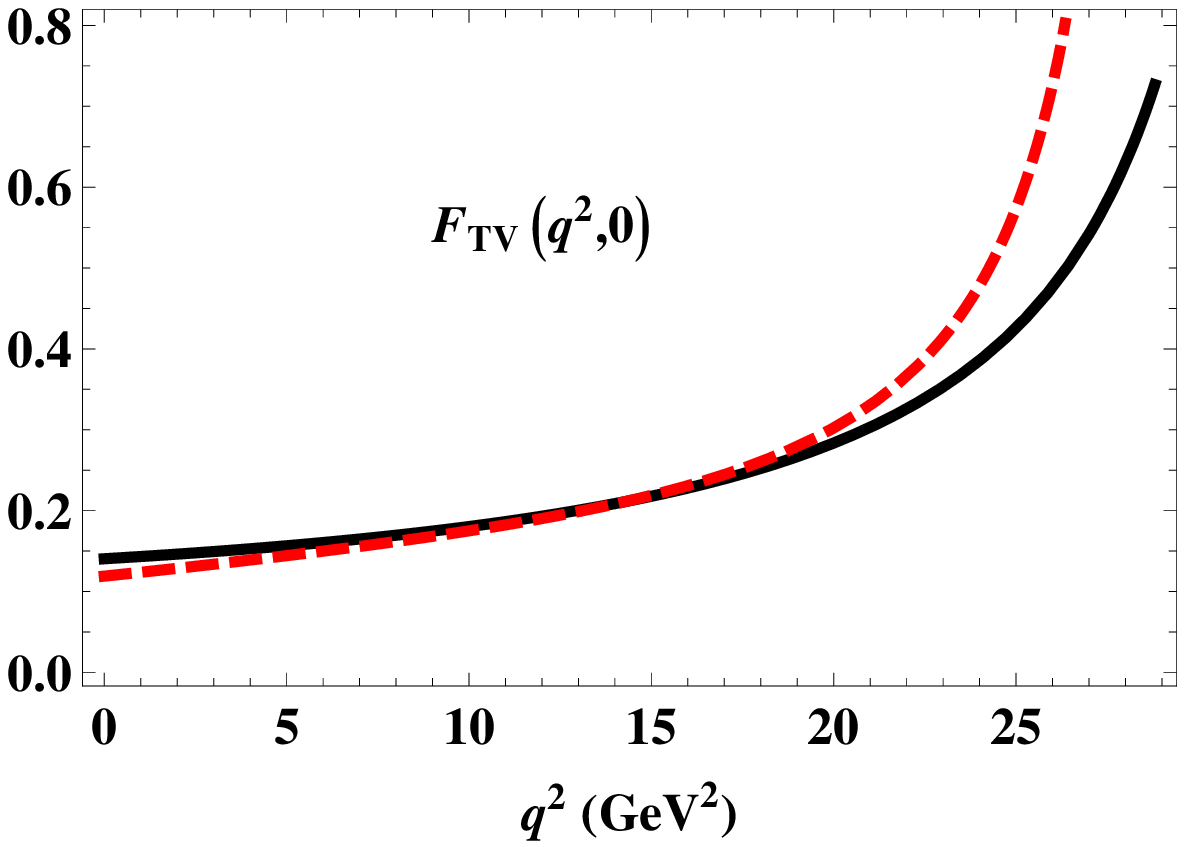}
\end{tabular}
\end{center}
\caption{\label{fig:ffMel}
Comparison of the form factors $F_i(q^2,0)$ calculated in our model (solid lines) with those from Ref.~\cite{Kozachuk:2017mdk} (dashed lines).
}
\end{figure}

In Figs.~\ref{fig:ffMel} we compare our form factors with the 
Kozachuk-Melikhov-Nikitin (KMN) form factors calculated in Ref.~\cite{Kozachuk:2017mdk}. Using the definitions in Eqs.~(\ref{eq:ffdef}) and~(\ref{eq:ffcont}) we can relate our form factors $F_i(q^2)$ to the KMN form factors $F_i(q^2,0)$ as follows 
(see Ref.~\cite{Kozachuk:2017mdk} for more detail):
\be
F_{V/A}(q^2,0) = F_{V/A}(q^2),\qquad F_{TV/TA}(q^2,0) = F_{TV/TA}(q^2)-e_b \tilde{F}_{TV/TA}^{b(\bar{l}l)b}-e_s \tilde{F}_{TV/TA}^{s(\bar{l}l)s}.
\en
One can see that in the low-$q^2$ region ($q^2\lesssim 20$ GeV$^2$) the corresponding form factors from the two sets are very close. In the high-$q^2$ region, the KMN form factors steeply increase and largely exceed our form factors.
\begin{figure}[ht]
\begin{center}
\begin{tabular}{ll} 
\includegraphics[scale=0.6]{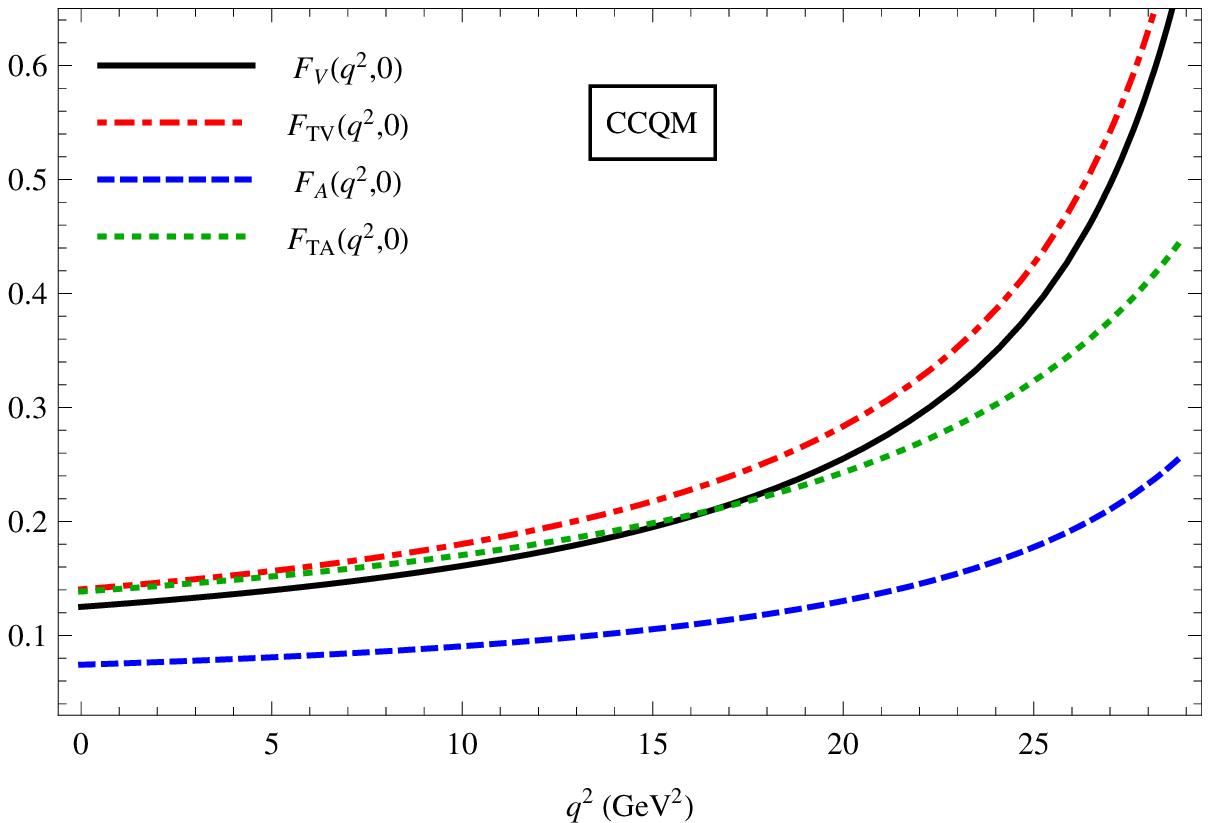} &
\includegraphics[scale=0.6]{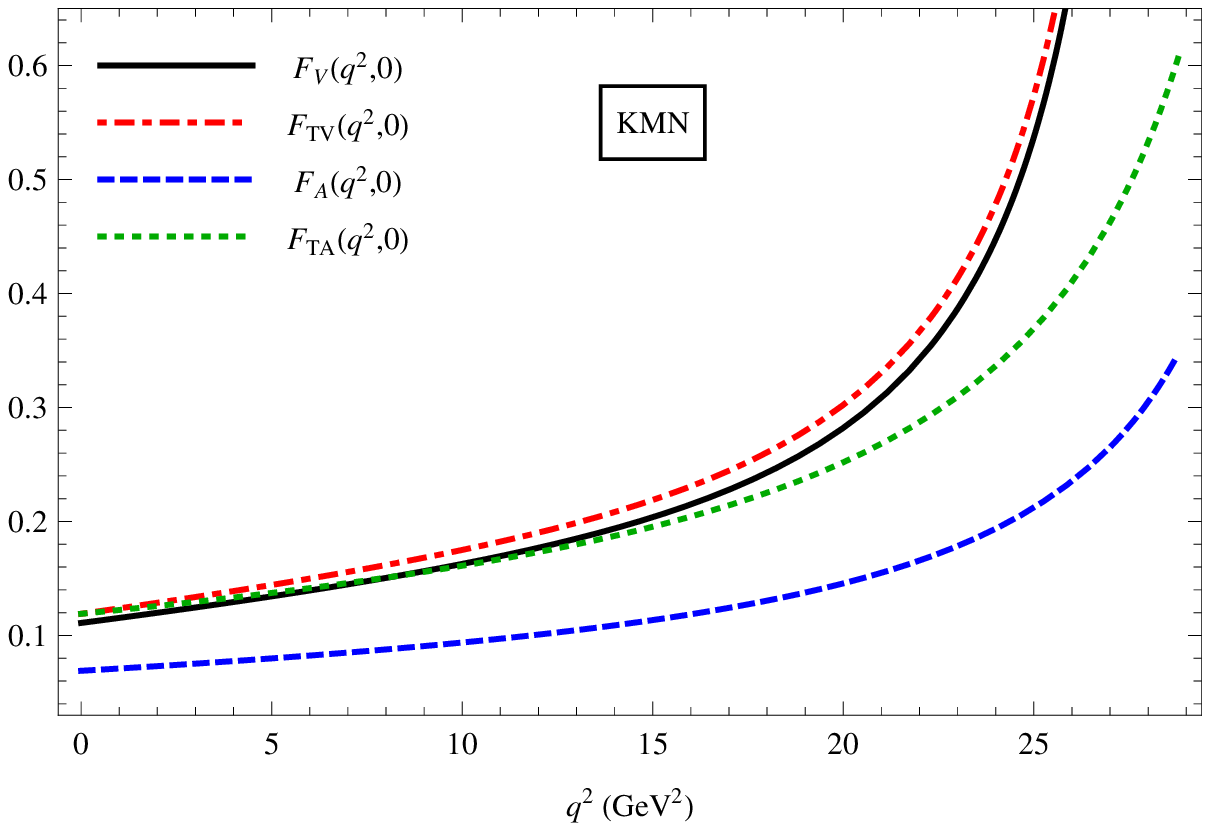}
\end{tabular}
\end{center}
\caption{\label{fig:allff}
Behavior of the form factors $F_i(q^2,0)$ in comparison with Ref.~\cite{Kozachuk:2017mdk} (KMN).
}
\end{figure}

In order to have a better picture of the behavior of the form factors, in Fig.~\ref{fig:allff} we plot all of them together and compare with those from Ref.~\cite{Kozachuk:2017mdk}. It is very interesting to note that our form factors share with the corresponding KMN ones not only similar shapes (especially in the low-$q^2$ region) but also relative behaviors, i.e., similar relations between the form factors, in the whole $q^2$ region. Several comments should be made:
(i) our form factors satisfy the constraint $F_{TA}(q^2,0)=F_{TV}(q^2,0)$ at $q^2=0$, with the common value equal to 0.135; (ii) in the small-$q^2$ region, $F_{V}(q^2,0)\approx F_{TA}(q^2,0)\approx F_{TV}(q^2,0)$; (iii) $F_{V}(q^2,0)$ and $F_{TV}(q^2,0)$ are approximately equal in the full kinematical range and rise steeply in the high-$q^2$ region; and (iv) $F_{A}(q^2,0)$ and $F_{TA}(q^2,0)$ are rather flat when $q^2\to M_{B_s}^2$ as compared to $F_{V}(q^2,0)$ and $F_{TV}(q^2,0)$. These observations show that our form factors satisfy very well the constraints on their behavior proposed by the authors of 
Ref.~\cite{Kruger:2002gf}.

\begin{figure}[htbp]
\begin{center}
\begin{tabular}{ll} 
\includegraphics[scale=0.55]{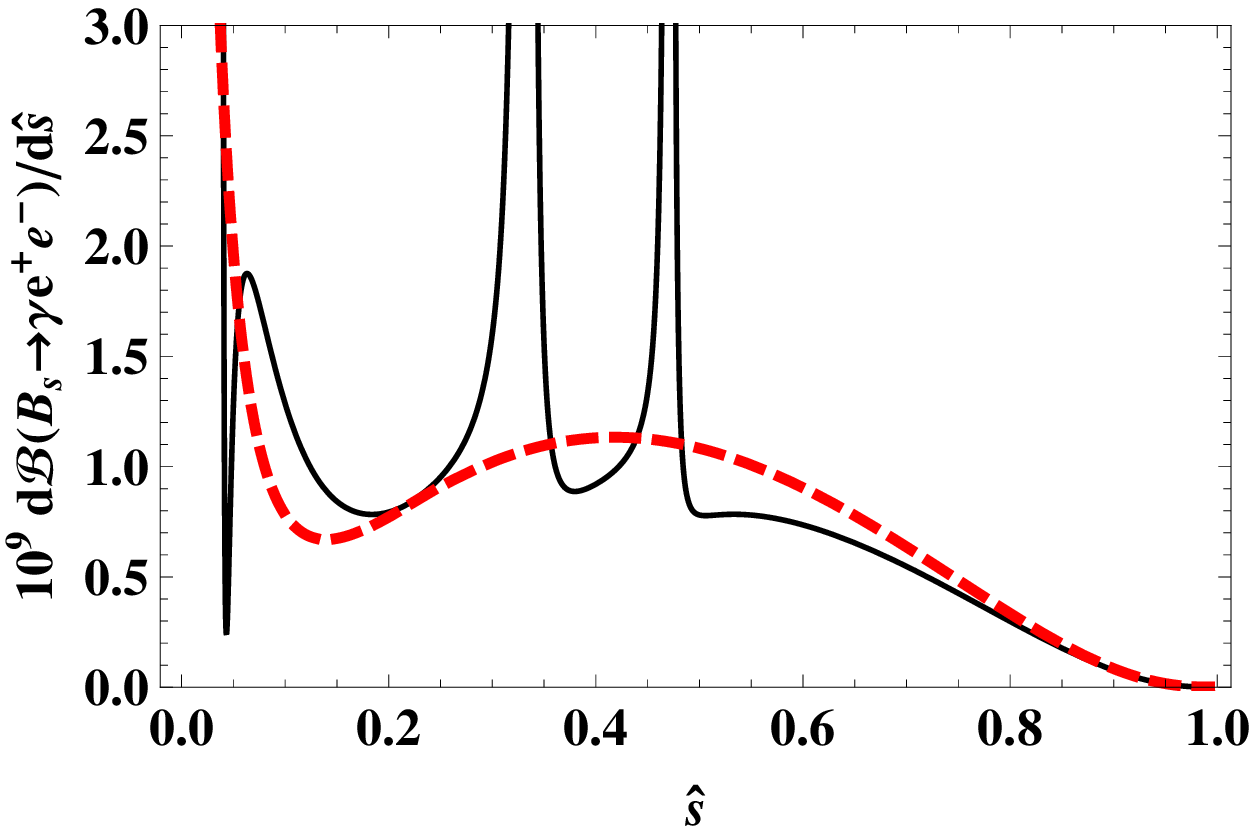} &
\includegraphics[scale=0.55]{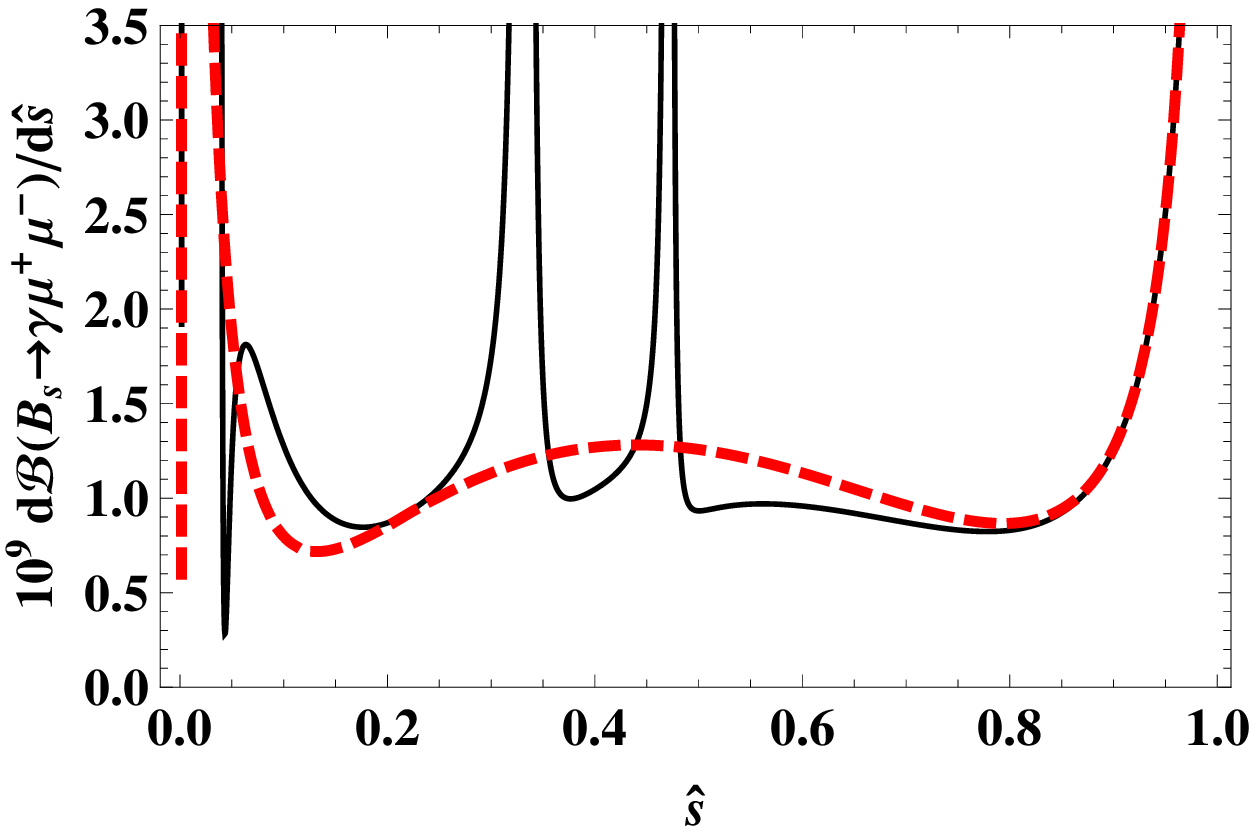} \\ 
\includegraphics[scale=0.55]{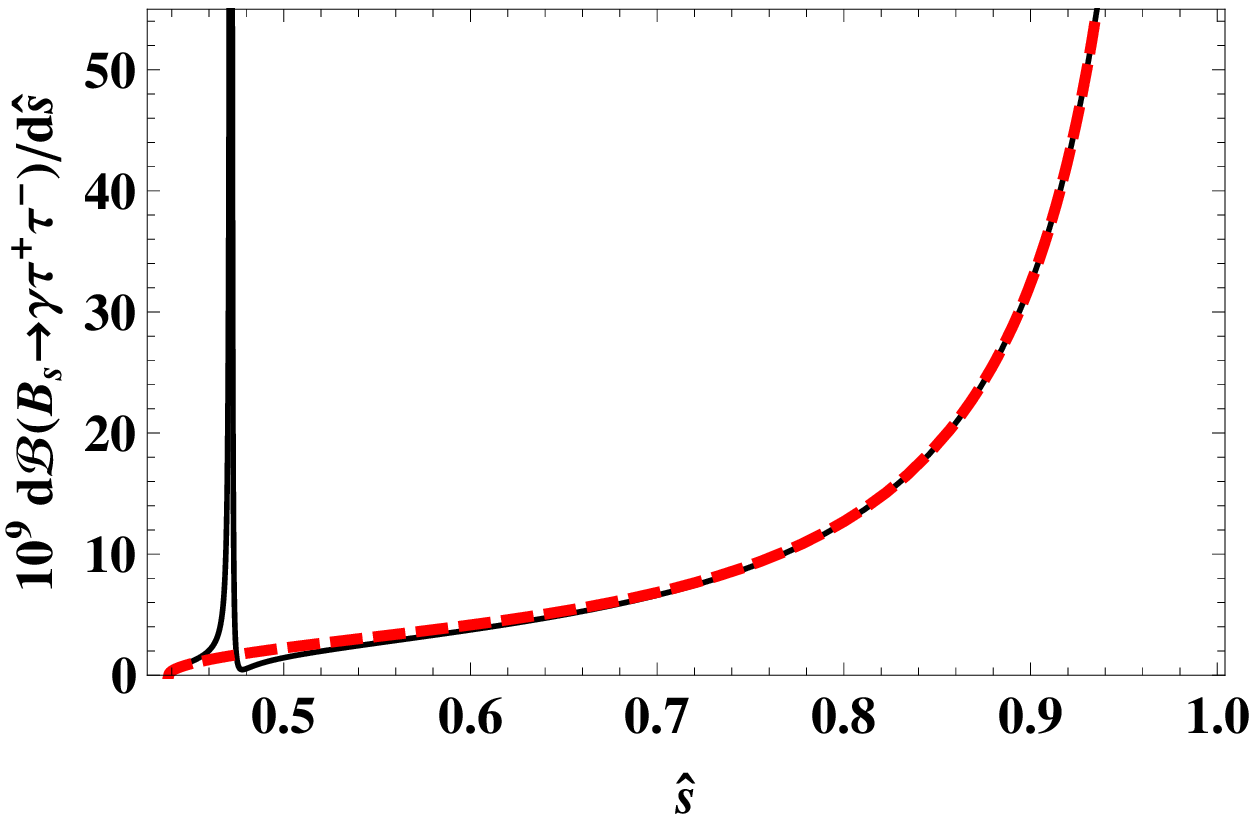} &
\includegraphics[scale=0.55]{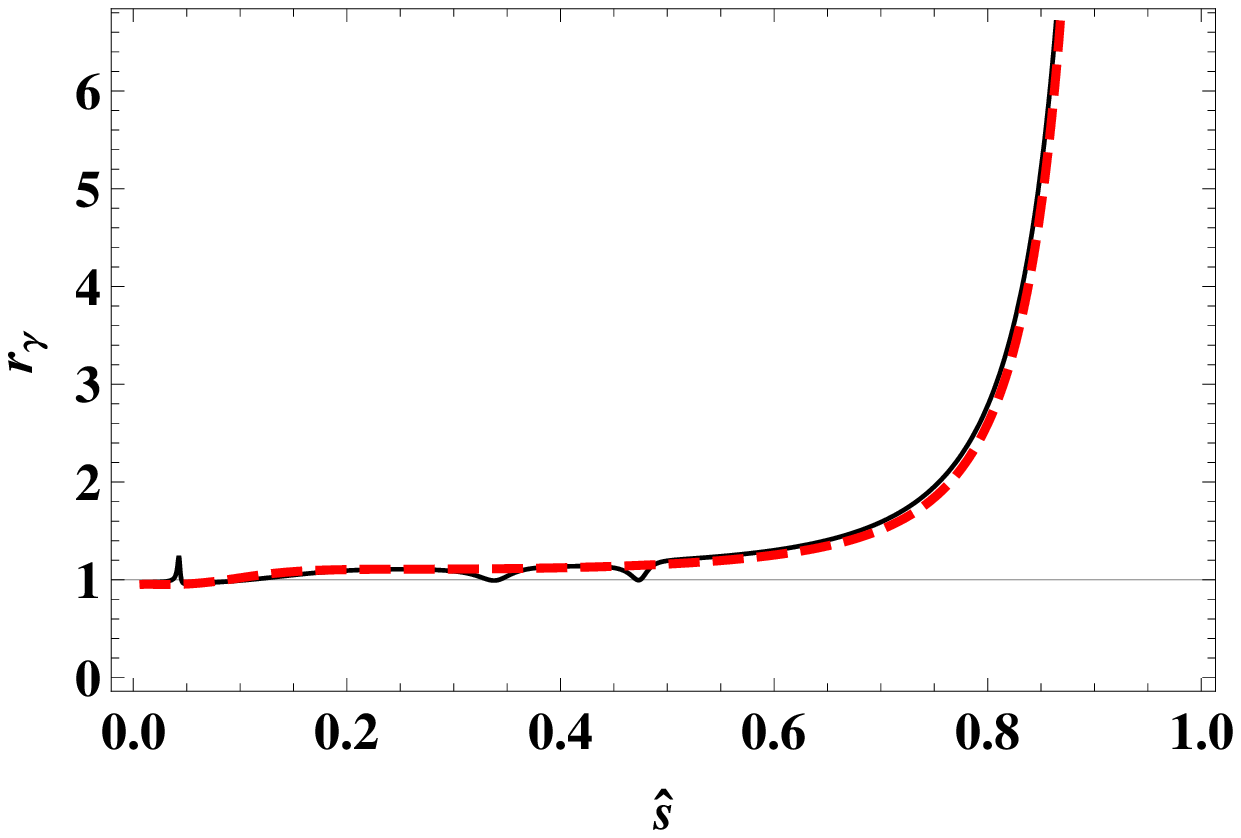}
\end{tabular}
\end{center}
\caption{\label{fig:Br}
Differential branching fractions 
$10^9d\mathcal{B}\left(B_s\to \gamma\ell^+\ell^-\right)/d\hat s$ and ratio $r_\gamma$
as functions of the dimensionless variable $\hat s = q^2/M^2_{B_s}$
without long-distance contributions (dashed lines) and with contributions
of the low lying charmonia $J/\psi$ and $\psi(2S)$, and the light $\phi$ meson (solid lines).
The photon energy cut $E_{\gamma\,\rm min}=20$~MeV is used.  
}
\end{figure}
\begin{figure}[t]
\includegraphics[scale=0.5]{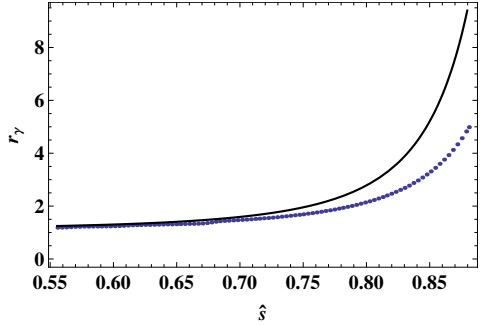}
\caption{\label{fig:Ratio}
Ratio $r_\gamma(\hat{s})$ at large $\hat{s}$ obtained in our model (solid line) and from 
Ref.~\cite{Kozachuk:2017mdk} (dotted line).}
\end{figure}

In Fig.~\ref{fig:Br} we plot the  differential branching fractions $10^9 d\mathcal{B}\left(B_s\to \gamma\ell^+\ell^-\right)/d\hat s$ 
as functions of the dimensionless variable $\hat s = q^2/M^2_{B_s}$. We also plot here the ratio 
\be
r_\gamma(\hat{s})\equiv \frac{d\mathcal{B}(B_s\to \gamma\mu^+\mu^-)/d\hat s}{{d\mathcal{B}(B_s\to \gamma e^+ e^-)/d\hat s}},
\en
which is a promising observable for testing lepton flavor universality (LFU) in these channels~\cite{Guadagnoli:2017quo}. The ratio $r_\gamma$ is very close to unity in the low-$q^2$ region but far above unity at large $q^2$ due to bremsstrahlung. As was pointed out in Ref.~\cite{Kozachuk:2017mdk}, in the high-$q^2$ region ($q^2\gtrsim 15$~GeV$^2$), the ratio $r_\gamma$ is mainly described by the form factors $F_A(q^2)$ and 
$F_V(q^2)$. Therefore, knowledge of their behavior at large $q^2$ plays an important role in testing LFU. In Fig.~\ref{fig:Ratio} we plot the ratio $r_\gamma$ at large $q^2$ in comparison with Ref.~\cite{Kozachuk:2017mdk}. The ratios are very close in the range
$16\lesssim q^2\lesssim 20$ GeV$^2$ (i.e., $0.55\lesssim\hat{s}\lesssim 0.69$), which is a  result of the similarity between the form factors discussed above.

The authors of Ref.~\cite{Guadagnoli:2017quo} suggested a useful observable
\be
R_\gamma(\hat{s}_1,\hat{s}_2)\equiv \frac{\int_{\hat{s}_1}^{\hat{s}_2}d\hat{s}\, d\mathcal{B}(B_s\to \gamma\mu^+\mu^-)/d\hat s}{{\int_{\hat{s}_1}^{\hat{s}_2}d\hat{s}\,d\mathcal{B}(B_s\to \gamma e^+ e^-)/d\hat s}},
\en
with the optimal choice $\hat{s}_1=0.55$ and $\hat{s}_2=0.8$ (corresponding to $q^2_1=15.8$ GeV$^2$ and $q^2_2=23.0$ GeV$^2$, respectively). In this range, the ratio is dominated by the form factors $F_A(q^2)$ and $F_V(q^2)$. We provide our prediction for this ratio $R_\gamma(0.55,0.8)=1.54$, about $30\%$ larger than the prediction $R_\gamma(0.55,0.8)=1.115\pm 0.030$ given in Ref.~\cite{Guadagnoli:2017quo}. Note that from the results of Ref.~\cite{Kozachuk:2017mdk}, one obtains 
$R_\gamma(0.55,0.8)=1.32$.

In Table~\ref{tab:br-1} we give the values of the branching fractions calculated without and with long-distance contributions. In the calculation with long-distance contributions, the region of two low lying charmonia is excluded by assuming $0.33\le \hat s\le 0.55$, as usually done in experimental data analysis. It is seen that the bremsstrahlung contribution to the electron mode is negligible, while for the tau mode it becomes the main part.

\begin{table*}[!htb]
\caption{Branching fractions with (in brackets) and without long-distance contributions.
The used minimal photon energy is $E_{\gamma\,\rm min}=20$~MeV.
}
\label{tab:br-1}
\begin{center}
\def\arraystretch{0.75}
\begin{ruledtabular}
\begin{tabular}{ccccc}
 {}                                & SD                  & BR           & IN                             & Sum \\
\hline
 $ 10^9\mathcal{B}(B_s\to \gamma e^+e^-) $       &  3.05 ({\bf 15.9})  & $3.2\times 10^{-5}$ &  $-4.8 ({\bf -9.5})\times  10^{-6}$  & 3.05 ({\bf 15.9})  \\
 $ 10^9 \mathcal{B}(B_s\to \gamma \mu^+\mu^-) $   &  1.16 ({\bf 10.0})  & 0.53              &  $-7.4 ({\bf -14.4})\times 10^{-3}$    & 1.7 ({\bf 10.5}) \\
 $ 10^9 \mathcal{B}(B_s\to \gamma \tau^+\tau^-) $ &  0.10 ({\bf 0.05})  & 13.4             &  0.30 ({\bf 0.18})            & 13.8 ({\bf 13.7}) \\
\end{tabular}
\end{ruledtabular}
\end{center}
\end{table*} 

In Table~\ref{tab:br-2} we compare our results for the branching fractions with those obtained in other approaches.  Our predictions for the electron and muon modes agree well with the results of Ref.~\cite{Melikhov:2017pwu}. In the case of the tau mode, the contribution from bremsstrahlung dominates the decay branching fractions and the SD amplitude becomes less important. As a result, our prediction for $\mathcal{B}(B_s\to \gamma \tau^+\tau^-)$ agrees well with other studies.
\begin{table}[ht]
\caption{Comparison of the branching fractions $10^9 \mathcal{B}(B_s\to \gamma \ell^+\ell^-)$ ($\ell=e,\mu,\tau$) with other approaches.
The used minimal photon energy is $E_{\gamma\,\rm min}=20$~MeV.
}
\label{tab:br-2}
\begin{center}
\def\arraystretch{0.75}
\begin{tabular}{cccc}
\hline
\hline
  Reference                 &\qquad Electron \qquad & \qquad  Muon\qquad      & \qquad  Tau \\
\hline
This work              &\qquad   15.9    \qquad & \qquad   10.5  \qquad    & \qquad 13.7 \\
\cite{Eilam:1996vg}    &\qquad   6.2    \qquad & \qquad   4.6  \qquad    &\qquad   $\dots$  \\
\cite{Aliev:1996ud}    &\qquad   2.35   \qquad & \qquad   1.9  \qquad    &\qquad   $\dots$  \\
\cite{Aliev:1997sd}    &\qquad   $\dots$     \qquad & \qquad   $\dots$    \qquad    &\qquad  15.2  \\
\cite{Geng:2000fs}     &\qquad  7.1     \qquad & \qquad   8.3  \qquad    &\qquad  15.7 \\
\cite{Dincer:2001hu}   &\qquad  20.0     \qquad & \qquad  12.0  \qquad    &\qquad  $\dots$ \\
\cite{Melikhov:2004mk} &\qquad  24.6    \qquad & \qquad   18.9 \qquad    &\qquad  11.6 \\
\cite{Melikhov:2017pwu} &\qquad  18.4    \qquad & \qquad   11.6 \qquad    &\qquad  $\dots$ \\
\cite{Wang:2013rfa}  &\qquad  17.4     \qquad & \qquad  17.4  \qquad    &\qquad $\dots$ \\
\hline
\hline
\end{tabular}
\end{center}
\end{table} 

Finally, in Tables~\ref{tab:br-bin} and~\ref{tab:br-bin2} we provide our predictions for the branching fractions integrated over several $q^2$ bins, which are more practical for experimental studies than the total branching fractions. We show also the corresponding results obtained by KMN~\cite{Kozachuk:2017mdk}, and Guadagnoli-Reboud-Zwicky (GRZ)~\cite{Guadagnoli:2017quo} for comparison. It is seen that our predictions agree quite well with both the KMN and GRZ results.
\begin{table}[ht]
\caption{Branching fractions $10^9\Delta\mathcal{B}(B_s\to \gamma \ell^+\ell^-)$ integrated in several $q^2$ bins. KMN results are given in brackets~\cite{Kozachuk:2017mdk}. Here, to obtain $q^2_{\rm max}$ we use $E_{\gamma\,{\rm min}}=80$ MeV  [see Eq.~(\ref{eq:Egam})], which was also used in Ref.~\cite{Kozachuk:2017mdk}, in order to define the same bin.}
\label{tab:br-bin}
\begin{center}
\def\arraystretch{0.8}
\begin{ruledtabular}
\begin{tabular}{cccccc}
 {}  & $\left[4m_e^2, 4m_\mu^2\right]$ & $\left[4m_\mu^2, 1\,{\rm GeV}^2\right]$  
 & $\left[1\,{\rm GeV}^2, 6\,{\rm GeV}^2\right]$         & $\left[6\,{\rm GeV}^2, 0.33M_{B_s}^2\right]$ &  $\left[0.55M_{B_s}^2, q^2_{\rm max}\right]$\\
\hline
 $B_s\to \gamma e^+e^-$       &  5.76 (4.67)  &   2.24 (1.80) &  7.50 (6.00) &  0.28 (0.14) & 0.17 (0.20)\\
$B_s\to \gamma \mu^+\mu^-$ &  $\dots$   &  2.05 (1.80)  & 7.50 (6.00)              &   0.29 (0.15) & 0.47 (0.43)   \\
\end{tabular}
\end{ruledtabular}
\end{center}
\end{table} 
\begin{table}[ht]
\caption{Branching fractions $10^9\Delta\mathcal{B}(B_s\to \gamma \mu^+\mu^-)$ integrated in low and high $q^2$ regions. Here, we use $E_{\gamma\,{\rm min}}=50$ MeV as in Ref.~\cite{Guadagnoli:2017quo}, in order to define the same bin.}
\label{tab:br-bin2}
\begin{center}
\def\arraystretch{0.8}
\begin{ruledtabular}
\begin{tabular}{ccc}
 {}  &  $\left[4m_\mu^2,\, 0.30M_{B_s}^2\right]$  
  &  $\left[0.55M_{B_s}^2,\, M_{B_s}^2-2E_{\gamma\,{\rm min}}M_{B_s}\right]$\\
\hline
 This work &  9.6   &  0.53  \\
 GRZ~\cite{Guadagnoli:2017quo} & $8.4\pm 1.3$ & $0.89\pm 0.10$
\end{tabular}
\end{ruledtabular}
\end{center}
\end{table} 

Our predictions for the branching fractions in Tables~\ref{tab:br-1}--\ref{tab:br-bin2} contain the uncertainties from the hadronic form factors and from other inputs, including the Wilson coefficients given in Table~\ref{tab:input}. However, the uncertainties from the latter are much smaller than those from the former. Therefore we estimate the errors of the branching fractions to be of order $30\%$ based on the uncertainty of the form factors. 

It is also important to note that the light $\phi$ meson resonance significantly enhances the branching fractions of the electron and muon modes. In the calculation above we have integrated over the whole $q^2$ range corresponding to the $\phi$ meson resonance. If we consider a small $q^2$ cut $[(m_\phi-\Gamma_\phi)^2, (m_\phi+\Gamma_\phi)^2]$ around the $\phi$ resonance, then we obtain
$10^9\Delta\mathcal{B}(B_s\to \gamma \ell^+\ell^-)=2.04$ for $q^2\in [1,6]\,{\rm GeV}^2$, where $\ell=e\,,\mu$.
\section{Summary and conclusions}
\label{sec:summary}
We have studied the rare radiative leptonic decays 
$B_s\to \gamma \ell^+\ell^-$ ($\ell=e,\mu,\tau$) in the framework of the covariant confined quark model. The relevant transition form factors have been obtained in the full kinematical range of dilepton momentum transfer squared. We have found a very good agreement between our form factors and those from Ref.~\cite{Kozachuk:2017mdk}, especially in the region $q^2\lesssim 20$ GeV$^2$. We have provided predictions for the decay branching fractions and their ratio. 
The branching fractions for the light leptons agree well with the results of Ref.~\cite{Melikhov:2017pwu}. For the tau mode, our prediction agrees well with all existing values in the literature. The branching fractions in different $q^2$ bins have also been calculated, and show good agreement with the results of Refs.~\cite{Kozachuk:2017mdk,Guadagnoli:2017quo}. In particular, we have predicted $10^9\Delta\mathcal{B}(B_s\to \gamma \ell^+\ell^-)=7.50$ for $q^2\in [1,6]\,{\rm GeV}^2$, where $\ell=e\,,\mu$.
\begin{acknowledgments}
The work was partly supported by Slovak Grant Agency for Sciences
(VEGA), Grant No. 1/0158/13 and Slovak Research and Development Agency (APVV), Grant No. APVV-0463-12 (S.~D., A.~Z.~D., A.~L.). S.~D., A.~Z.~D., M.~A.~I, and A.~L. acknowledge the support from the Joint Research Project of Institute of Physics, Slovak Academy of Sciences (SAS), and Bogoliubov Laboratory of Theoretical Physics, Joint Institute for Nuclear Research (JINR), Grant No. 01-3-1114. P.~S. acknowledges the support from Istituto Nazionale di Fisica Nucleare, 
I.S. QFT\_\,HEP. C.~T.~T. thanks Dmitri Melikhov for useful discussions.
\end{acknowledgments}


\end{document}